\documentclass[a4paper,10pt]{revtex4}
\usepackage{mathrsfs}
\usepackage{graphicx}
\usepackage{latexsym}
\usepackage{amsmath}
\usepackage{amssymb}
\usepackage{textcomp}
\usepackage{amsbsy}
\usepackage{graphics}
\usepackage{epstopdf}
\usepackage{color}

\allowdisplaybreaks[4]
\begin{document}

\title{Different faces of generalized holographic dark energy}

\author{Shin'ichi~Nojiri$^{1,2}$\,\thanks{nojiri@gravity.phys.nagoya-u.ac.jp},
S.~D.~Odintsov$^{3,4}$\,\thanks{odintsov@ieec.uab.es},
Tanmoy~Paul$^{5,6}$\,\thanks{pul.tnmy9@gmail.com}} \affiliation{
$^{1)}$ Department of Physics, Nagoya University,
Nagoya 464-8602, Japan \\
$^{2)}$ Kobayashi-Maskawa Institute for the Origin of Particles
and the Universe, Nagoya University, Nagoya 464-8602, Japan \\
$^{3)}$ ICREA, Passeig Luis Companys, 23, 08010 Barcelona, Spain\\
$^{4)}$ Institute of Space Sciences (IEEC-CSIC) C. Can Magrans
s/n, 08193 Barcelona, Spain\\
$^{5)}$ Department of Physics, Chandernagore College, Hooghly - 712 136.\\
$^{6)}$ International Laboratory for Theoretical Cosmology, TUSUR, 634050 Tomsk, Russia}

\begin{abstract}
In the formalism of generalized holographic dark energy (HDE), the holographic 
cut-off is generalized to depend upon $L_\mathrm{IR} = L_\mathrm{IR} \left( 
L_\mathrm{p}, \dot L_\mathrm{p}, 
\ddot L_\mathrm{p}, \cdots, L_\mathrm{f}, \dot L_\mathrm{f}, \cdots, a\right)$ with $L_\mathrm{p}$ and $L_\mathrm{f}$ 
are the particle horizon and the future horizon, respectively (moreover 
$a$ is the scale factor of the universe). Based on such formalism, in the present paper, 
we show that a wide class of dark energy (DE) models can be regarded as different candidates of the generalized HDE 
family, with respective cut-offs. This can be thought as a symmetry between the generalized HDE and different DE models. 
In this regard, we consider several entropic dark energy models - like Tsallis entropic DE, the 
R\'{e}nyi entropic DE, and the Sharma-Mittal entropic DE - and showed that 
they are indeed equivalent with the generalized HDE. Such equivalence between 
the entropic DE and the generalized HDE is extended to the scenario 
where the respective exponents of the entropy functions are allowed to 
vary with the expansion of the universe. Besides the entropic DE models, 
the correspondence with the generalized HDE is also established for the Quintessence and for the 
Ricci DE models. In all the above cases, the effective equation of state (EoS) parameter corresponds to the holographic energy density 
are determined, by which the equivalence of various DE models with the respective generalized HDE models are further confirmed. 
The equivalent holographic cut-offs are determined 
by two ways: (1) in terms of the particle horizon and its 
derivatives, (2) in terms of the future horizon horizon and its derivatives. 
\end{abstract}

\maketitle

\section{Introduction}

The holographic principle originates from black hole thermodynamics and string theory and establishes a connection 
of the infrared cutoff of a quantum field theory, which is related to the vacuum energy, with the largest distance of 
this theory \cite{tHooft:1993dmi,Susskind:1994vu,Witten:1998qj,Bousso:2002ju}. 
Such holographic consideration is extensively applied in the field cosmology, particular in describing the dark energy (DE) 
era, generally known as holographic dark energy (HDE) model 
\cite{Li:2004rb,Li:2011sd,Wang:2016och,Pavon:2005yx,Nojiri:2005pu,Enqvist:2004xv,Zhang:2005yz,Guberina:2005fb,Elizalde:2005ju,
Ito:2004qi,Gong:2004cb,Saridakis:2007cy,Gong:2009dc,BouhmadiLopez:2011xi,Malekjani:2012bw,Khurshudyan:2014axa,Khurshudyan:2016gmb,Landim:2015hqa,
Gao:2007ep,Li:2008zq,Anagnostopoulos:2020ctz,Zhang:2005hs,Li:2009bn,Feng:2007wn,Zhang:2009un,Lu:2009iv,
Micheletti:2009jy,Huang:2004wt,Mukherjee:2017oom,Nojiri:2017opc,Sharif:2019seo,Jawad:2018juh}. 
It may be stressed that instead of adding an extra term in the Lagrangian, the HDE 
is based on the holographic principle and on the dimensional analysis, and from this perspective, the HDE is significantly different 
than the other dark energy models. Apart from the dark energy era, the holographic principle is also successfully 
applied to the early inflationary universe 
\cite{Horvat:2011wr,Nojiri:2019kkp,Paul:2019hys,Bargach:2019pst,Elizalde:2019jmh,Oliveros:2019rnq}. 
Actually during the early universe the size of the universe was small, due to which, the 
holographic energy density becomes significant to trigger an inflationary scenario, and moreover the holographic inflation is also found to be compatible 
with the 2018 Planck constraints. Recently, some of our authors showed that the energy density coming from the holographic principle is able to 
unify the early inflationary scenario with the late dark energy era in a covariant formalism \cite{Nojiri:2020wmh}. From a different viewpoint, 
the application of the holographic principle is extended to the bouncing scenario in \cite{Nojiri:2019yzg,Brevik:2019mah,Coriano:2019eif,Elizalde:2020zcb,
Odintsov:2020zct}, where the holographic energy density 
helps to violate the energy conditions and in turn leads to a bouncing universe.

Coming back to the dark energy context, the holographic dark energy density 
is proportional to the inverse squared of the holographic cut-off ($L_\mathrm{IR}$) which is usually assumed to be same as the particle horizon 
($L_\mathrm{p}$) or the future horizon ($L_\mathrm{f}$). However the fundamental form of the $L_\mathrm{IR}$ is still a debatable topic in this context. 
Along this line, it deserves mentioning that the most generalized cut-off has been proposed in \cite{Nojiri:2005pu}, where in particular, 
the cut-off is considered to depend upon $L_\mathrm{IR} = L_\mathrm{IR}(L_\mathrm{p}, \dot L_\mathrm{p}, 
\ddot L_\mathrm{p}, \cdots, L_\mathrm{f}, \dot L_\mathrm{f}, \cdots, a)$, 
which in turn leads to the generalized version of HDE (known as ``generalized HDE''). 
Such generalized form of $L_\mathrm{IR}$ immediately leads to the following question:

\begin{itemize}
\item Does there exist suitable form(s) of $L_\mathrm{IR}$ such that various dark energy models (including the entropic DE models) 
can be thought to be equivalent to the generalized HDE? If so, then what will be the equivalent 
form(s) of $L_\mathrm{IR}$ for the respective DE models?
\end{itemize}

In the present paper, we will intend to address the above questions. For this purpose, we will consider several entropic DE models, like - the Tsallis 
entropic DE \cite{Tsallis:1987eu,Lyra:1998wz,Wilk:1999dr,Tsallis:2012js,Komatsu:2013qia,Nunes:2014jra,Lymperis:2018iuz,Saridakis:2018unr, 
Sheykhi:2018dpn,Artymowski:2018pyg,Abreu:2017hiy,Jawad:2018frc,Zadeh:2018wub, 
daSilva:2018ehn}, the R\'{e}nyi entropic DE \cite{Biro:2013cra,Czinner:2015eyk,Komatsu:2016vof,Moradpour:2017ycq,Moradpour:2018ima,Moradpour:2018ivi} 
and the Sharma-Mittal entropic DE \cite{Jahromi:2018xxh,Masi,Moradpour:2018ima} respectively. Here it may be mentioned that 
unlike to the Tsallis and the R\'{e}nyi entropy functions, the Sharma-Mittal entropy represents a more general two parameter functions. 
In this regard, our investigation will be carried for two 
cases: (1) where the respective exponents of the entropy functions are treated as constant, while in the second case, (2) the exponents 
are allowed to vary with cosmic time, in particular for the latter case, the exponents are considered to depend on the evolving Hubble parameter 
of the universe. Besides the entropic DE models, the Quintessence \cite{Halliwell:1986ja,Barreiro:1999zs,Rubano:2001su,
Sangwan:2018zpz} and the Ricci-DE \cite{Gao:2007ep,Zhang:2009un,delCampo:2013hka,Granda:2008dk} 
models will also take part in the present analysis. In case 
of Quintessence model, a non-minimally coupled scalar field with an exponential potential provides the dark energy density, while in the Ricci-DE scenario, 
the space-time Ricci curvature serves the dark energy density. Both the Quintessence and the Ricci-DE models turn out to be viable with respect to 
various dark energy observations. Interestingly, we will show that 
all such entropic DE, Quintessence and the Ricci-DE models are indeed equivalent with the 
generalized HDE, with suitable forms of the corresponding cut-offs.

\section{Thermodynamics of Space-Time and Application to Cosmology}\label{Sec0}

After the thermodynamical properties of the black hole were clarified 
\cite{Bekenstein:1974ax,Hawking:1974sw} and it has been claimed that the entropy of the 
black hole is proportional to the area $A$ of the horizon 
\begin{equation}
\label{Tslls5}
S = \frac{A}{4G}\, ,\quad A = 4\pi r_H^2\, ,
\end{equation}
which is called as the Bekenstein-Hawking entropy,  
$r_H$ is the horizon radius, and we work in units where $\hbar=k_B = c = 1$, 
there have been long and active studies where the connection between the gravity 
and the thermodynamics 
could be clarified \cite{Jacobson:1995ab,Padmanabhan:2003gd,Padmanabhan:2009vy}. 
In the studies, we have found that the FRW equations can be also 
regarded as the first law of 
thermodynamics when we consider the Bekenstein-Hawking entropy by using 
the cosmological apparent horizon \cite{Cai:2005ra,Akbar:2006kj,Cai:2006rs} 
as a realization of the thermodynamics of space-time \cite{Jacobson:1995ab}. 

In case that, however, there are long range forces like the electro-magnetic one or 
gravitational one, we know that the systems are non-additive systems and 
the standard Boltzmann-Gibbs additive entropy should not be applied and we should 
generalize the entropy to the non-extensive Tsallis entropy 
\cite{Tsallis:1987eu,Lyra:1998wz,Wilk:1999dr} and recently there are several attempts in this regard 
(see \cite{Tsallis:2012js,Komatsu:2013qia,Nunes:2014jra,Lymperis:2018iuz,Saridakis:2018unr, 
Sheykhi:2018dpn,Artymowski:2018pyg,Abreu:2017hiy,Jawad:2018frc,Zadeh:2018wub, 
daSilva:2018ehn}). 
If we apply the Tsallis entropy to the black hole, instead of the 
Bekenstein-Hawking entropy, one finds \cite{Tsallis:2012js},
\begin{equation}
\label{Tslls9}
S_\mathrm{T} = \frac{A_0}{4 G} \left(\frac{A}{A_0} \right)^\delta\, .
\end{equation}
In the above expression, $A_0$ is a constant and $\delta$ 
is the new parameter that quantifies the non-extensivity. 
Then if we apply the Tsallis entropy by using the apparent horizon to the 
cosmology, the FRW equations should be modified and the modification can be regarded 
as the contribution from the dark energy. 

In information theory, the R\'{e}nyi entropy is often used as the measure of 
the entanglement. 
If we apply the R\'{e}nyi entropy to the black hole, one finds 
\cite{Biro:2013cra,Czinner:2015eyk,Komatsu:2016vof,Moradpour:2017ycq,Moradpour:2018ima}
\begin{equation}
\label{Re1}
S_\mathrm{R} = \frac{A_0}{G \delta} \ln \left( 1 + \frac{\delta}{4}\left(\frac{A}{A_0}\right) \right) \, .
\end{equation}
The R\'{e}nyi entropy has been also used to explain the dark energy \cite{Moradpour:2018ima}. 

Here it may be mentioned that both the 
Tsallis and R\'{e}nyi entropy expressions belong from one-parametric entropy family; there is also 
a two-parametric generalized entropy which is called the Sharma-Mittal entropy ($S_\mathrm{SM}$) and is written as \cite{Jahromi:2018xxh,Masi,Moradpour:2018ima},
\begin{eqnarray}
S_\mathrm{SM} = \frac{A_0}{G\alpha}\left\{\left( 
1 + \frac{\delta}{4}\left(\frac{A}{A_0} \right)\right)^{\frac{\alpha}{\delta}} - 1\right\} \, ,
 \label{SM entropy}
\end{eqnarray}
where $A_0$ is a constant, $\alpha$ and $\delta$ are two independent parameters. 
Some cosmic applications of the Sharma-Mittal entropy can be found in \cite{Jahromi:2018xxh} 
where the Hubble horizon plays the role of cut-off and moreover no mutual interaction 
between the cosmos components has been taken into account.

Above considerations of different entropies eventually lead to different scenarios of holographic dark energy, which will be discussed in the following 
two sections.

\section{Dark Energy corresponding to Tsallis, R\'{e}nyi, and Sharma-Mittal entropies} \label{SecI}


We assume the Friedmann-Lema\^{i}tre-Robertson-Walker (FLRW) space-time with flat spacial part, whose 
metric is given by 
\begin{equation}
ds^2=-dt^2+a^2(t)\sum_{i=1,2,3} \left(dx^i\right)^2 \, .
\label{metric}
\end{equation}
Here $a(t)$ is called as a scale factor. 

If we define the Hubble rate $H$ by $H=\frac{\dot a}{a}$, the radius $r_H$ of the 
cosmological horizon is given by 
\begin{equation}
\label{apphor}
r_H=\frac{1}{H}\, .
\end{equation}
Then the entropy in the region inside the cosmological horizon could be given by 
the Bekenstein-Hawking relation \cite{Padmanabhan:2009vy} in (\ref{Tslls5}). 
On the other hand, the flux of the energy $E$ or the increase of the heat $Q$ in the 
region is given by 
\begin{equation}
\label{Tslls2}
dQ = - dE = -\frac{4\pi}{3} r_H^3 \dot\rho dt = -\frac{4\pi}{3H^3} \dot\rho dt 
= \frac{4\pi}{H^2} \left( \rho + p \right) dt \, ,
\end{equation}
where we use the conservation law: $0 = \dot \rho + 3 H \left( \rho + p \right)$. 
Then by using the Hawking temperature 
\cite{Cai:2005ra}
\begin{equation}
\label{Tslls6}
T = \frac{1}{2\pi r_H} = \frac{H}{2\pi}\, ,
\end{equation}
and the first law of thermodynamics $TdS = dQ$, 
one obtains $\dot H = - 4\pi G \left( \rho + p \right)$ 
and by integrating the expression , one obtains 
the first FLRW equation, 
\begin{equation}
\label{Tslls8}
H^2 = \frac{8\pi G}{3} \rho + \frac{\Lambda}{3} \, .
\end{equation}
Here the cosmological constant $\Lambda$ appears as a constant of the integration. 

Instead of the Bekenstein-Hawking entropy (\ref{Tslls5}), we may 
use the non-extensive, the Tsallis entropy
\cite{Tsallis:1987eu,Lyra:1998wz,Wilk:1999dr,Nunes:2014jra} 
in (\ref{Tslls9}). 
Then by applying the first law of thermodynamics to the system, 
instead of $\dot{H} = - 4\pi G \left( \rho + p \right)$, one gets \cite{Lymperis:2018iuz}
\begin{eqnarray}
 \delta\left(\frac{H^2}{H_1^2}\right)^{1-\delta}\dot{H} = - 4\pi G \left( \rho + p \right)\, ,
 \label{Tsallis-new1}
\end{eqnarray}
on integrating which, one gets,
\begin{equation}
\label{Tslls11}
\frac{\delta}{2 - \delta} H_1^2\left( \frac{H^2}{H_1^2} \right)^{2 - \delta}
= \frac{8\pi G}{3} \rho + \frac{\Lambda}{3} \, .
\end{equation}
Here a constant $H_1$ is defined by $A_0 \equiv \frac{4\pi}{H_1^2}$. 
Then if we define the energy density $\rho_\mathrm{T}$ and the pressure $p_\mathrm{T}$ by 
\begin{eqnarray}
\rho_\mathrm{T} = \frac{3}{8\pi G} 
\left( H^2 - \frac{\delta}{2 - \delta} H_1^2\left( \frac{H^2}{H_1^2} \right)^{2 - \delta} 
\right)\, ,\label{rhoT}
\end{eqnarray}
\begin{eqnarray}
p_\mathrm{T} = \frac{\dot{H}}{4\pi G}\left\{\delta\left(\frac{H^2}{H_1^2}\right)^{1-\delta} - 1\right\} 
- \frac{3}{8\pi G} 
\left( H^2 - \frac{\delta}{2 - \delta} H_1^2\left( \frac{H^2}{H_1^2} \right)^{2 - \delta} 
\right)~,
\label{pT}
\end{eqnarray}
respectively. It is evident that $\rho_\mathrm{T}$ depends on the quadratic power of the Hubble parameter and thus is symmetric with respect to the 
Hubble parameter. With the above forms of $\rho_\mathrm{T}$, $p_\mathrm{T}$, Eqs.~(\ref{Tsallis-new1}) and (\ref{Tslls11}) can be expressed as 
\begin{eqnarray}
\dot{H}&=&-4\pi G\left[\left(\rho + p\right) + \left(\rho_\mathrm{T} + p_\mathrm{T}\right)\right]~, \nonumber\\
H^2&=&\frac{8\pi G}{3} \left( \rho_\mathrm{T} + \rho \right) 
+ \frac{\Lambda}{3} \, ,
\label{Tslls11BB}
\end{eqnarray}
respectively. Therefore $\rho_\mathrm{T}$ and $p_\mathrm{T}$ represent the energy density and pressure correspond 
to Tsallis entropy. Consequently the respective equation of state (EoS) parameter for the Tsallis entropy is given by,
\begin{eqnarray}
 \omega_\mathrm{T} = \frac{p_\mathrm{T}}{\rho_\mathrm{T}} 
 = -1 + 2\left(\frac{\dot{H}}{H^2}\right)\left\{\frac{\delta\left(\frac{H^2}{H_1^2}\right)^{1-\delta} - 1}
 {1 - \frac{\delta}{2 - \delta} \left( \frac{H^2}{H_1^2} \right)^{1 - \delta}}\right\}
 \label{eosT}
\end{eqnarray}
It may be checked that the above expression of $\omega_\mathrm{T}$ leads to the conservation equation for the Tsallis entropic energy density, i.e
\begin{eqnarray}
 \dot{\rho}_\mathrm{T} + 3H\rho_\mathrm{T}\left(1 + \omega_\mathrm{T}\right) = 0~~.
 \label{conservation-T}
\end{eqnarray}
Here it deserves mentioning that the authors of \cite{Lymperis:2018iuz} showed that the $\omega_\mathrm{T}$ in Eq.(\ref{eosT}) leads to a viable 
dark energy epoch of our present universe, where the matter sector is considered to be dust. Moreover, the analysis is also extended 
to the case where the radiation energy density is present too. In particular, due to the Tsallis entropic energy density, 
the universe exhibits the usual thermal history, with the sequence of matter and dark-energy eras and the onset of acceleration occurs at 
around $z \approx 0.5$ which is in agreement with observations \cite{Lymperis:2018iuz}.

In regard to the R\'{e}nyi entropy (\ref{Re1}), the first law of thermodynamics gives, 
\begin{equation}
\label{RFRW01}
 - \frac{ H^3 \dot H}
{H^2 + \frac{\delta}{4} H_1^2} 
= - \frac{4\pi G}{3} \dot\rho \, , 
\end{equation}
from which, we obtain 
\begin{equation}
\label{RFRW}
H^2 = \frac{8\pi G}{3} \rho + \frac{\Lambda}{3} 
+ \frac{\delta}{4} H_1^2 \ln \left( \frac{H^2}{H_1^2} +  \frac{\delta}{4} \right) \, .
\end{equation}
Here the cosmological constant $\Lambda$ appears as a constant of the integration again. 
At this stage we may define 
the corresponding energy density and the pressure in the following form 
\begin{eqnarray}
\rho_\mathrm{R} = \frac{3\delta}{32G} H_1^2 
\ln \left( \frac{H^2}{H_1^2} +  \frac{\delta}{4} \right) \, ,\label{Re2}
\end{eqnarray}
\begin{eqnarray}
p_\mathrm{R} = -\frac{\dot{H}}{4\pi G}\left\{\frac{1}{1 + \frac{4}{\delta}\left(\frac{H^2}{H_1^2}\right)}\right\} 
- \frac{3\delta}{32G} H_1^2 
\ln \left( \frac{H^2}{H_1^2} +  \frac{\delta}{4} \right)~~.
\label{pR}
\end{eqnarray}
Similar to the Tsallis entropic case, the R\'{e}nyi entropic 
energy density (i.e $\rho_\mathrm{R}$) seems to be symmetric in respect to the Hubble parameter. 
Due to the above expressions of $\rho_\mathrm{R}$ and 
$p_\mathrm{R}$, Eqs.(\ref{RFRW01}) and (\ref{RFRW}) become similar to the usual Friedmann equations 
where the total energy density and total pressure are given by $\rho_\mathrm{eff} = \rho + \rho_\mathrm{R}$ and $p_\mathrm{eff} = p + p_\mathrm{R}$. 
Consequently, the EoS parameter corresponds to the R\'{e}nyi entropy comes with the following form,
\begin{eqnarray}
 \omega_\mathrm{R} = \frac{p_\mathrm{R}}{\rho_\mathrm{R}} 
 = -1 - \frac{8}{3\pi\delta}\left(\frac{\dot{H}}{H_1^2}\right)\left\{\frac{1}{\ln \left( \frac{H^2}{H_1^2} +  \frac{\delta}{4} \right)
 \left[1 + \frac{4}{\delta}\left(\frac{H^2}{H_1^2}\right)\right]}\right\}~.
 \label{eosR}
\end{eqnarray}
It may be mentioned that the above expression of $\omega_\mathrm{R}$ obeys the conservation equation for the R\'{e}nyi entropic energy density. 
As showed in \cite{Moradpour:2018ima,Moradpour:2018ivi}, the R\'{e}nyi entropic energy density ($\rho_\mathrm{R}$) and the pressure ($p_\mathrm{R}$) 
can provide suitable description for the current accelerated universe and thus leads to a dark energy model.

In case of the Sharma-Mittal entropy, the first law of thermodynamics leads to the following evolution of the cosmic Hubble parameter,
\begin{eqnarray}
\left(1 + \frac{\delta H_1^2}{4H^2} \right)^{\frac{\alpha}{\delta} - 1} \dot{H} 
= -4\pi G \left( \rho + p \right)\, ,
 \label{SM1}
\end{eqnarray}
integrating which, we obtain,
\begin{eqnarray}
H_1^2\left( \frac{\left(\frac{\delta}{4} \right)^{\frac{\alpha}{\delta} - 1}}{2 - \alpha/\delta}\right) 
\left( \frac{H^2}{H_1^2} \right)^{2 - \frac{\alpha}{\delta}} 
{}_2F_1\left[1 - \frac{\alpha}{\delta}, 2 - \frac{\alpha}{\delta}, 3 - \frac{\alpha}{\delta}; -\frac{4}{\delta} 
\left( \frac{H^2}{H_1^2} \right) \right] 
= \frac{8\pi G}{3} \rho + \frac{\Lambda}{3}~,
\label{SM2}
\end{eqnarray}
where $\Lambda$ is the constant of integration, 
${}_2F_1$ is the hypergeometric function, and to get the above expression, 
we use the conservation equation of the matter components. Moreover, 
the constant $H_1$ is related to $A_0$ by $A_0 = \frac{4\pi}{H_1^2}$. Now if we define an energy density ($\rho_\mathrm{SM}$) 
and a pressure ($p_\mathrm{SM}$) like,
\begin{eqnarray}
\rho_\mathrm{SM} = \frac{3}{8\pi G}
\left\{H^2 - H_1^2\left(\frac{\left(\frac{\delta}{4} \right)^{\frac{\alpha}{\delta} - 1}}{2 - \frac{\alpha}{\delta}} \right) 
\left(\frac{H^2}{H_1^2} \right)^{2 - \frac{\alpha}{\delta}} 
{}_2F_1 \left[ 1 - \frac{\alpha}{\delta}, 2 - \frac{\alpha}{\delta}, 3 - \frac{\alpha}{\delta}; 
 -\frac{4}{\delta} \left( \frac{H^2}{H_1^2} \right) \right] \right\} \, ,\label{SM3}
 \end{eqnarray}
\begin{eqnarray}
 p_\mathrm{SM} = \frac{\dot{H}}{4\pi G}\left\{\left(1 + \frac{\delta H_1^2}{4H^2} \right)^{\frac{\alpha}{\delta} - 1} - 1\right\} - \rho_\mathrm{SM}\, ,
 \label{pSM}
\end{eqnarray}
respectively, then Eqs.~(\ref{SM1}) and (\ref{SM2}) can be equivalently expressed as,
\begin{eqnarray}
\dot{H}&=&-4\pi G\left[\left(\rho + p\right) + \left(\rho_\mathrm{SM} + p_\mathrm{SM}\right)\right]~, \nonumber\\
 H^2&=&\frac{8\pi G}{3} \left( \rho_\mathrm{SM} + \rho \right) 
+ \frac{\Lambda}{3}\, .
\label{SM4}
\end{eqnarray}
Thus we may argue that $\rho_\mathrm{SM}$ and $p_\mathrm{SM}$ are the energy density and the pressure coming from the cosmological description 
of the Sharma-Mittal entropy. Furthermore $\rho_\mathrm{SM}$ and $p_\mathrm{SM}$ are connected by the respective EoS, as given by
\begin{eqnarray}
 \omega_\mathrm{SM} = -1 + \left(\frac{\dot{H}}{3H^2}\right)
 \left\{\frac{\left(1 + \frac{\delta H_1^2}{4H^2} \right)^{\frac{\alpha}{\delta} - 1} - 1}
 {1 - \left(\frac{\left(\frac{\delta}{4} \right)^{\frac{\alpha}{\delta} - 1}}{2 - \frac{\alpha}{\delta}} \right) 
\left(\frac{H_1^2}{H^2} \right)^{\frac{\alpha}{\delta}-1} 
{}_2F_1 \left[ 1 - \frac{\alpha}{\delta}, 2 - \frac{\alpha}{\delta}, 3 - \frac{\alpha}{\delta}; 
 -\frac{4}{\delta} \left( \frac{H^2}{H_1^2} \right) \right]}\right\}~~,
 \label{eosSM}
\end{eqnarray}
where we use Eqs.(\ref{SM3}) and (\ref{pSM}). The above form of $\omega_\mathrm{SM}$ immediately confirms the conservation equation for the 
Sharma-Mittal entropic energy density. Furthermore, as established in \cite{Jahromi:2018xxh,Moradpour:2018ima}, 
the Sharma-Mittal entropic energy density leads to a late time acceleration epoch of our universe. In \cite{Jahromi:2018xxh}, the 
universe in considered to be filled by a pressureless component and Sharma-Mittal entropic energy density, which do not have 
any mutual interaction, and as a result the present deceleration parameter is found to be consistent with the present observation.\\

Before closing this section, here we would like to mention that the presence of entropic energy densities indeed modify the FLRW equations. Such 
modifications can also be encapsulated in the respective entropy functions. In particular, when we consider the Bekenstein-Hawking entropy in the context of 
cosmology, one gets the usual FLRW equations and thus we have the expression like $\frac{dS}{dH} = -2\pi/\left(GH^3\right)$ 
which leads to $S = S(H)$. However for the modified entropy cases, the relation of $S = S(H)$ will become different compared to the Bekenstein-Hawking 
case. In particular, for the Tsallis entropy case, the first FLRW Eq.(\ref{Tslls11}) leads to the following expression:
\begin{eqnarray}
 \delta\left(\frac{H^2}{H_1^2}\right)^{1-\delta}HdH = \frac{4\pi G}{3}d\rho~,
 \label{new-1}
\end{eqnarray}
which along with the first law of thermodynamics (in Eq.(\ref{Tslls2})) yield the rspective entropy in terms of the Hubble parameter as,
\begin{eqnarray}
 \frac{dS_\mathrm{T}}{dH} = -\frac{2\pi \delta}{G}\left(\frac{1}{H^3}\right)\left(\frac{H^2}{H_1^2}\right)^{1-\delta}~,
 \label{new-2}
\end{eqnarray}
on integrating which, one gets $S_\mathrm{T} = S_\mathrm{T}(H)$. 
It is evident that for $\delta = 1$, the above expression becomes similar to that of the Bekenstein-Hawking case. Thereby the modification of the 
Tsallis entropy compared to the Bekenstein-hawking case is clearly demonstrated by the expression of $S_\mathrm{T} = S_\mathrm{T}(H)$. 
By similar procedure, we can obtain the R\'{e}nyi entropy and the Sharma-Mittal entropy functions in terms of the Hubble parameter as,
\begin{eqnarray}
 \frac{dS_\mathrm{R}}{dH}&=&-\frac{2\pi}{G}\left(\frac{1}{H^3}\right)\frac{1}{\left(1 + \frac{\delta}{4}\left(\frac{H_1^2}{H^2}\right)\right)}\label{new-3}~,\\
 \frac{dS_\mathrm{SM}}{dH}&=&-\frac{2\pi}{G}\left(\frac{1}{H^3}\right)\left(1 + \frac{\delta}{4}\left(\frac{H_1^2}{H^2}\right)\right)
 ^{\frac{\alpha}{\delta}-1}~~.
 \label{new-4}
\end{eqnarray}
Clearly Eq.(\ref{new-3}) depicts that for $\delta = 1$, one gets $\frac{dS_\mathrm{R}}{dH} = -2\pi/\left(GH^3\right)$, while Eq.(\ref{new-4}) reveals that 
the situation $\alpha = \delta$ leads to $\frac{dS_\mathrm{SM}}{dH}=-2\pi/\left(GH^3\right)$, i.e they become similar with that of the Bekenstein-Hawking 
entropy function for the aforesaid conditions respectively. Here we would like to mention that 
Eqs.(\ref{new-2}), (\ref{new-3}) and (\ref{new-4}) remains symmetric under the transformation $H \rightarrow -H$.

\section{Generalized Holographic Energy}\label{SecII}

In the holographic principle, the holographic energy
density is proportional to the inverse squared infrared cutoff
$L_\mathrm{IR}$, which could be related with the causality given
by the cosmological horizon,
\begin{equation}
\label{basic}
\rho_\mathrm{hol}=\frac{3c^2}{\kappa^2 L^2_\mathrm{IR}}\, .
\end{equation}
Here $\kappa^2=8\pi G$ is the gravitational constant and $c$ is a free
parameter. 
The infrared cutoff $L_\mathrm{IR}$ is usually assumed to be the
particle horizon $L_\mathrm{p}$ or the future event horizon
$L_\mathrm{f}$, which are given as,
\begin{equation}
\label{H3}
L_\mathrm{p}\equiv a \int_0^t\frac{dt}{a}\ ,\quad
L_\mathrm{f}\equiv a \int_t^\infty \frac{dt}{a}\, .
\end{equation}
Differentiating both sides of the above expressions lead to the Hubble parameter in terms of $L_\mathrm{p}$, $\dot{L}_\mathrm{p}$ or in terms of 
$L_\mathrm{f}$, $\dot{L}_\mathrm{f}$ as,
\begin{equation}
\label{HLL}
H \left( L_\mathrm{p} , \dot{L}_\mathrm{p} \right) =
\frac{\dot{L}_\mathrm{p}}{L_\mathrm{p}} - \frac{1}{L_\mathrm{p}}\, , 
\quad 
H(L_\mathrm{f} , \dot{L}_\mathrm{f}) = \frac{\dot{L}_\mathrm{f}}{L_\mathrm{f}} + \frac{1}{L_\mathrm{f}}
\, .
\end{equation}
In \cite{Nojiri:2005pu}, a general form of the cutoff was proposed,
\begin{equation}
\label{GeneralLIR}
L_\mathrm{IR} = L_\mathrm{IR} \left( L_\mathrm{p}, \dot L_\mathrm{p}, 
\ddot L_\mathrm{p}, \cdots, L_\mathrm{f}, \dot L_\mathrm{f}, \cdots, a\right)\, .
\end{equation}
Actually, the other dependency of $L_\mathrm{IR}$, particularly on the Hubble parameter, Ricci scalar and their derivatives, can be 
transformed to either $L_p$ and their derivatives or $L_f$ and their derivatives via Eq.(\ref{HLL}).
The above cutoff could be chosen to be equivalent to a general covariant gravity model,
\begin{equation}
\label{GeneralAc}
S = \int d^4 \sqrt{-g} F \left( R,R_{\mu\nu} R^{\mu\nu},
R_{\mu\nu\rho\sigma}R^{\mu\nu\rho\sigma}, \Box R, \Box^{-1} R,
\nabla_\mu R \nabla^\mu R, \cdots \right) \, .
\end{equation}
We will use the above expressions frequently in the following sections. 
With the help of the generalized cut-off, we aim to show that the Tsallis, R\'{e}nyi and Sharma-Mittal entropic dark energy 
may belong from the generalized dark energy family where the holographic cut-offs are expressed in terms of the particle horizon and its derivatives or 
in terms of the future horizon and its derivatives.

The comparison of Eqs.~(\ref{rhoT}) and (\ref{basic}) lead to the argument that the Tsallis entropic dark energy belongs 
from the generalized holographic dark energy family, where the corresponding infrared cutoff $L_\mathrm{T}$ is given by, 
\begin{align}
\label{rhoT2}
\frac{3c^2}{\kappa^2 L^2_\mathrm{T}} 
= \frac{3}{8\pi G} 
\left( \left( \frac{\dot{L}_\mathrm{p}}{L_\mathrm{p}} - \frac{1}{L_\mathrm{p}} \right)^2 
- \frac{\delta}{2 - \delta} H_1^2\left( \frac{\left( \frac{\dot{L}_\mathrm{p}}{L_\mathrm{p}} - \frac{1}{L_\mathrm{p}} \right)^2}
{H_1^2} \right)^{2 - \delta} \right) \, , 
\end{align}
in terms of $L_\mathrm{p}$ and its derivatives. To get the above expression, we use Eq.~(\ref{HLL}). Moreover, $L_\mathrm{T}$ in terms of the future horizon and its 
derivatives comes by the following way,
\begin{align}
\label{rhoTfuture horizon}
\frac{3c^2}{\kappa^2 L^2_\mathrm{T}} = \frac{3}{8\pi G} 
\left( \left( \frac{\dot{L}_\mathrm{f}}{L_\mathrm{f}} + \frac{1}{L_\mathrm{f}} \right)^2 
- \frac{\delta}{2 - \delta} H_1^2\left( \frac{\left( \frac{\dot{L}_\mathrm{f}}{L_\mathrm{f}} + \frac{1}{L_\mathrm{f}} \right)^2}
{H_1^2} \right)^{2 - \delta} \right) \, .
\end{align}
Here we would like to determine the EoS parameter of the holographic energy density corresponds to the cut-off $L_\mathrm{T}$, 
in particular of $\rho_\mathrm{hol} = 3c^2/\left(\kappa^2L_\mathrm{T}^2\right)$. In this regard, 
the conservation equation of $\rho_\mathrm{hol}$ immediately yields the respective EoS parameter (symbolized by $\Omega_\mathrm{hol}^{(T)}$) as,
\begin{eqnarray}
 \Omega_\mathrm{hol}^{(T)} = -1 - \left(\frac{2}{3HL_\mathrm{T}}\right)\frac{dL_\mathrm{T}}{dt}~,
 \label{eos-holT}
\end{eqnarray}
 where $L_\mathrm{T}$ is obtained in Eq.(\ref{rhoT2}) (or Eq.(\ref{rhoTfuture horizon})) and the superscript 'T' 
 in the above expression denotes the EoS parameter corresponds to the holographic cut-off $L_\mathrm{T}$. Due to 
 Eq.(\ref{HLL}), the above form of $\Omega_\mathrm{hol}^{(T)}$ seems to be 
 equivalent to the EoS of the Tsallis entropic energy density presented in Eq.(\ref{eosT}), i.e $\Omega_\mathrm{hol}^{(T)} \equiv \omega_\mathrm{T}$. 
 Such equivalence, along with the fact that the Tsallis entropic energy density provides a viable dark energy model, lead to the argument 
 that the holographic energy density coming from the cut-off $L_\mathrm{T}$ is also able to produce a viable dark energy epoch at our current universe. 

 Similarly, by comparing (\ref{Re2}) and (\ref{basic}),
the infrared cutoff $L_\mathrm{R}$ corresponding to the R\'{e}nyi entropy is 
given by 
\begin{equation}
\label{Re3}
\frac{3c^2}{\kappa^2 L^2_\mathrm{R}}
= \frac{3\delta}{32G} H_1^2 
\ln \left( \frac{1}{H_1^2} 
\left( \frac{\dot{L}_\mathrm{p}}{L_\mathrm{p}} - \frac{1}{L_\mathrm{p}} \right)^2 +  \frac{\delta}{4} 
\right) 
= \frac{3\delta}{32G} H_1^2 
\ln \left( \frac{1}{H_1^2} 
\left( \frac{\dot{L}_\mathrm{f}}{L_\mathrm{f}} + \frac{1}{L_\mathrm{f}} \right)^2 +  \frac{\delta}{4} 
\right) \, ,
\end{equation}
where, once again, we use Eq.~(\ref{HLL}). The first expression of Eq.~(\ref{Re3}) gives the $L_\mathrm{R}$ in terms of $L_\mathrm{p}$ and 
its derivatives, while the second one represents the same in terms of $L_\mathrm{f}$ and its derivatives. Once again, the conservation 
equation of the holographic energy density $\rho_\mathrm{hol} = 3c^2/\left(\kappa^2L_\mathrm{R}^2\right)$ leads to the corresponding EoS parameter 
($\Omega_\mathrm{hol}^{(R)}$) as,
\begin{eqnarray}
 \Omega_\mathrm{hol}^{(R)} = -1 - \left(\frac{2}{3HL_\mathrm{R}}\right)\frac{dL_\mathrm{R}}{dt}~,
 \label{eos-holR}
\end{eqnarray}
where $L_\mathrm{R}$ is given in Eq.(\ref{Re3}). It can be easily checked that the $\Omega_\mathrm{hol}^{(R)}$ satisfies the 
conservation relation: $\dot{\rho}_\mathrm{hol} + 3H\left(1 + \Omega_\mathrm{hol}^{(R)}\right) = 0$, where $\rho_\mathrm{hol}$ represents the 
holographic energy density coming from the cut-off $L_\mathrm{R}$. Eqs.(\ref{HLL}) and (\ref{Re3}) 
indicate that the above expression of $\Omega_\mathrm{hol}^{(R)}$ proves to be equivalent with $\omega_\mathrm{R}$ in 
Eq.(\ref{eosR}), i.e $\Omega_\mathrm{hol}^{(R)} \equiv \omega_\mathrm{R}$. Thereby, since the R\'{e}nyi entropic energy density 
suitably describes the current acceleration of our universe, we may argue that the holographic energy density coming from 
$L_\mathrm{R}$ is able to produce the late time cosmic acceleration.

Finally Eqs.~(\ref{SM3}) and (\ref{basic}) clearly argue that the Sharma-Mittal entropic dark energy can also be thought as one of the 
candidates of the generalized dark energy family, where the corresponding cut-off ($L_\mathrm{SM}$) is given by,
\begin{align}
\frac{3c^2}{\kappa^2 L_\mathrm{SM}^2}=&
\frac{3}{8\pi G}\left\{\left( \frac{\dot{L}_\mathrm{p}}{L_\mathrm{p}} - \frac{1}{L_\mathrm{p}} \right)^2 
 - H_1^2\left(\frac{\left( \frac{\delta}{4} \right)^{\frac{\alpha}{\delta} - 1}}{2 - \alpha/\delta} \right)
\left(\frac{\left( \frac{\dot{L}_\mathrm{p}}{L_\mathrm{p}} - \frac{1}{L_\mathrm{p}} \right)^2}{H_1^2}\right)^{2 - \frac{\alpha}{\delta}} 
\right. \nonumber\\
& \left. \qquad \qquad \times{}_2F_1 \left[1 - \frac{\alpha}{\delta}, 2 - \frac{\alpha}{\delta}, 3 - \frac{\alpha}{\delta}; 
 -\frac{4}{\delta}\left(\frac{\left( \frac{\dot{L}_\mathrm{p}}{L_\mathrm{p}} - \frac{1}{L_\mathrm{p}} \right)^2}{H_1^2}\right)\right] \right\}\, ,
 \label{GHDE_SM1}
\end{align}
in terms of the particle horizon and its derivatives. Similarly, the $L_\mathrm{SM}$ in terms of the future horizon and its derivatives is given by,
\begin{align}
\frac{3c^2}{\kappa^2 L_\mathrm{SM}^2}=&\frac{3}{8\pi G}\left\{\left( \frac{\dot{L}_\mathrm{f}}{L_\mathrm{f}} + \frac{1}{L_\mathrm{f}} \right)^2 
 - H_1^2\left(\frac{\left( \frac{\delta}{4} \right)^{\frac{\alpha}{\delta} - 1}}{2 - \frac{\alpha}{\delta}} \right)
\left(\frac{\left( \frac{\dot{L}_\mathrm{f}}{L_\mathrm{f}} + \frac{1}{L_\mathrm{f}} \right)^2}{H_1^2}\right)^{2 - \frac{\alpha}{\delta}} \right. \nonumber\\
& \left. \qquad \qquad \times{}_2F_1\left[1 - \frac{\alpha}{\delta}, 2 - \frac{\alpha}{\delta}, 3 - \frac{\alpha}{\delta}; 
 -\frac{4}{\delta}\left(\frac{\left( \frac{\dot{L}_\mathrm{f}}{L_\mathrm{f}} + \frac{1}{L_\mathrm{f}} \right)^2}{H_1^2}\right)\right] \right\}\, .
\label{GHDE_SM2}
\end{align}
Furthermore using the conservation relation of $\rho_\mathrm{hol} = 3c^2/\left(\kappa^2L_\mathrm{SM}^2\right)$, we determine 
the EoS parameter ($\Omega_\mathrm{hol}^{(SM)}$) corresponds to the holographic energy density coming from the cut-off $L_\mathrm{SM}$ as,
\begin{eqnarray}
 \Omega_\mathrm{hol}^{(SM)} = -1 - \left(\frac{2}{3HL_\mathrm{SM}}\right)\frac{dL_\mathrm{SM}}{dt}~,
 \label{eos-holSM}
\end{eqnarray}
where $L_\mathrm{SM}$ is given in Eq.(\ref{GHDE_SM1}) (or in Eq.(\ref{GHDE_SM2})). In effect of Eq.(\ref{HLL}), it is evident that 
the above form of $\Omega_\mathrm{hol}^{(SM)}$ is equivalent to the EoS of the Sharma-Mittal entropic energy density of Eq.(\ref{eosSM}), i.e 
$\Omega_\mathrm{hol}^{(SM)} = \omega_\mathrm{SM}$. Due to this equivalence, we may argue that the holographic energy density 
$\rho_\mathrm{hol} = 3c^2/\left(\kappa^2L_\mathrm{SM}^2\right)$ can produce the late time acceleration of our universe. 

Therefore the dark energy models coming from the Tsallis entropy, the R\'{e}nyi entropy and 
the Sharma-Mittal entropy can be thought as different candidates of the generalized holographic dark energy family, where the respective 
infrared cutoffs are given by Eq.~(\ref{rhoT2}) to Eq.~(\ref{GHDE_SM2}) respectively. Thereby such holographic cut-offs establish 
a symmetry between generalized HDE and the respective entropic DE model(s). 


\section{Extended cases of entropic dark energy models} \label{Sec4}

In this section we consider the models extended as in \cite{Nojiri:2019itp}, where the 
non-extensive exponent  $\delta$ of the Tsallis entropy 
(\ref{Tslls9}) or the R\'{e}nyi entropy (\ref{Re1}) 
depends on the energy scale and shows a running behavior
\cite{Nojiri:2019skr}. 
In \cite{Nojiri:2019skr}, it has been claimed that such behaviors may appear because 
the entropy corresponds to physical degrees of freedom and 
the degrees of freedom depend on the scale as implied by 
the renormalization of a quantum theory. 
In case of gravity,  if the space-time  fluctuates at high energy scales, 
the degrees of freedom may increase. 
On the other hand, if gravity becomes a topological theory, the 
degrees of freedom may decrease. 

In cosmology, if we assume that the energy scale could be given by the 
Hubble scale $H$, $\delta$ in (\ref{Tslls9}) or (\ref{Re1}) may depend on $H$ 
\cite{Nojiri:2019skr}. Thereby we use a dimensionless variable $x\equiv \frac{H_1^2}{H^2} $ by introducing a parameter 
$H_1$ whose dimension is identical with $H$.
Then in case of the Tsallis entropy (\ref{Tslls9}), 
instead of (\ref{Tslls11}), one obtains the following generalized first FLRW equation (see the appendix for the detailed derivation), 
\begin{equation}
\label{Tslls17}
\left. - H_1^2 \left\{ x^{\delta(x) - 2} + 2 \int^x dx x^{\delta(x) -3} 
\right\}
\right|_{x=\frac{H_1^2}{H^2}}= \frac{8\pi G}{3} \rho + \frac{\Lambda}{3} \, ,
\end{equation}
and we may define the effective energy density $\rho_\mathrm{T}$ by 
\begin{equation}
\label{rhoT2C}
\rho_\mathrm{T} \equiv \frac{3}{8\pi G} 
\left( H^2 + \left. H_1^2 \left\{ x^{\delta(x) - 2} + 2 \int^x dx x^{\delta(x) -3} 
\right\}\right|_{x=\frac{H_1^2}{H^2}} \right)\, .
\end{equation}
Thereby the presence of a varying exponent in the Tsallis entropy modifies the FLRW equations, which may have considerable impacts in the universe's 
evolution, both at high and low energy scales. 
Clearly, in order to determine the explicit expression of $\rho_\mathrm{T}$ from the above equation, one needs a functional form of $\delta(x)$. 
Some of our authors proposed a suitable form of $\delta(x)$ in \cite{Nojiri:2019skr}, which allows to analytically perform 
the integration in Eq.(\ref{rhoT2C}) 
and also leads to an unified scenario of early inflation with late time acceleration. Actually, the form of $\delta(x)$ is chosen in such a way that 
at high and low energy scales it acquires values away from the standard value $1$, while at intermediate scales it comes close to unity 
\cite{Nojiri:2019skr}.

In case of the R\'{e}nyi entropy (\ref{Re1}),  
the first law of thermodynamics gives (see the appendix for the detailed derivation), 
\begin{equation}
\label{RFRW0001}
H^2 = \frac{8\pi G}{3} \rho + \frac{\Lambda}{3} 
+ 4 H_1^2 \left. \int^x dx x^{-2} \left\{ \left(\frac{\frac{\delta(x)}{16}}
{1 + \frac{\delta(x)}{4} x}\right)
+ \left( \frac{1}{\delta(x)^2} \ln \left( 1 + \frac{\delta(x)}{4} x \right) 
 - \frac{1}{\delta(x)} \frac{\frac{1}{4} x }
{1 + \frac{\delta(x)}{4} x} \right) 
\delta' (x) \right\}\right|_{x=\frac{H_1^2}{H^2}} \, ,
\end{equation}
where $\delta'(x) = \frac{d\delta}{dx}$. Then we may define the effective energy density $\rho_\mathrm{R}$ 
corresponding to R\'{e}nyi entropy by 
\begin{equation}
\label{rhoR2}
\rho_\mathrm{R} \equiv 4 H_1^2 \left. \int^x dx x^{-2} \left\{ \left(\frac{\frac{\delta(x)}{16}}
{1 + \frac{\delta(x)}{4} x}\right)
+ \left( \frac{1}{\delta(x)^2} \ln \left( 1 + \frac{\delta(x)}{4} x \right) 
 - \frac{1}{\delta(x)} \frac{\frac{1}{4} x }
{1 + \frac{\delta(x)}{4} x} \right) 
\delta' (x) \right\}\right|_{x=\frac{H_1^2}{H^2}} \, .
\end{equation}
The terms containing $\delta'(x)$ in the above expression of $\rho_\mathrm{R}$ arise due to the varying exponent $\delta = \delta(x)$. Such terms 
may play an important role during the early as well as in the late cosmic evolution of the universe. In particular, we expect that at high and low 
energy scales, the $\delta(x)$ should deviate from the standard value unity and thus have a significant role in driving the inflation or the late 
dark energy epoch. However the modified cosmology from the R\'{e}nyi entropy with a varying exponent has not been extensively studied in various 
earlier literatures. Thereby it will be an interesting avenue to study the possible effects of $\rho_\mathrm{R}$ in Eq.(\ref{rhoR2}) in the 
context of inflation, late time acceleration or even in the bouncing scenario. However these are out of the scope from the present work and thus 
we expect to study it in a future work.

Coming back to the Sharma-Mittal entropy, there are two independent parameters ($\alpha$ and $\delta$) and in the extended scenario, we take 
$\alpha = \alpha(x)$ and $\delta$ to be constant. However, in the extended case of the Sharma-Mittal entropy, 
one may choose both the parameters being dependent on $x = H_1^2/H^2$, i.e., $\alpha = \alpha(x)$ and $\delta = \delta(x)$. For simplicity, here we 
stick to the aforementioned consideration, i.e., $\alpha = \alpha(x)$ and $\delta = \mathrm{constant}$. Such consideration leads to the FLRW 
equation as (see the appendix for the detailed derivation),
\begin{eqnarray}
\left. -H_1^2 f(x) \right|_{x=\frac{H_1^2}{H^2}} = \frac{8\pi G}{3} \rho + \frac{\Lambda}{3}\, ,
 \label{SM-extended1}
\end{eqnarray}
with,
\begin{eqnarray}
f(x) = \int^{x}dx x^{-2}\left\{\left(1 + \frac{\delta x}{4}\right)^{\frac{\alpha(x)}{\delta}-1} - \frac{4\alpha'(x)}{\alpha(x)}
\left[\left(1 + \frac{\delta x}{4}\right)^{\frac{\alpha(x)}{\delta}} - 1\right] 
+\frac{4\alpha'(x)}{\alpha(x)\delta}\left(1 + \frac{\delta x}{4}\right)^{\frac{\alpha(x)}{\delta}}
\ln{\left(1 + \frac{\delta x}{4}\right)}\right\} \, .
\label{f}
\end{eqnarray}
Consequently, the effective energy density corresponds to the Sharma-Mittal entropy comes by
\begin{eqnarray}
\rho_\mathrm{SM} = \left. \frac{3}{8\pi G}\left\{H^2 + H_1^2 f(x)\right\} \right|_{x=\frac{H_1^2}{H^2}}\, .
\label{SM-extended2}
\end{eqnarray}
Eqs.~(\ref{rhoT2C}), (\ref{rhoR2}), and (\ref{SM-extended2}) immediately tell that the infrared cut-off $L_\mathrm{T}$, $L_\mathrm{R}$ and $L_\mathrm{SM}$ 
corresponding to the extended version of the Tsallis entropy, the R\'{e}nyi entropy, and the Sharma-Mittal entropy are given by, 
\begin{align}
\label{rhoT2B}
\frac{3c^2}{\kappa^2 L^2_\mathrm{T}} 
=& \frac{3}{8\pi G} 
\left( \left( \frac{\dot{L}_\mathrm{p}}{L_\mathrm{p}} - \frac{1}{L_\mathrm{p}} \right)^{2} + \left. H_1^2 \left\{ x^{\delta(x) - 2} + 2 \int^x dx x^{\delta(x) -3} 
\right\}\right|_{x=H_1^2\left( \frac{\dot{L}_\mathrm{p}}{L_\mathrm{p}} - \frac{1}{L_\mathrm{p}} \right)^{-2}} \right)\, \nonumber\\
=& \frac{3}{8\pi G} 
\left( \left( \frac{\dot{L}_\mathrm{f}}{L_\mathrm{f}} + \frac{1}{L_\mathrm{f}} \right)^{2} + \left. H_1^2 \left\{ x^{\delta(x) - 2} + 2 \int^x dx x^{\delta(x) -3} 
\right\}\right|_{x=H_1^2\left( \frac{\dot{L}_\mathrm{f}}{L_\mathrm{f}} + \frac{1}{L_\mathrm{f}} \right)^{-2}} \right)\, ,
\end{align}
\begin{align}
\label{Re3B}
\frac{3c^2}{\kappa^2 L^2_\mathrm{R}}
=& \frac{4 H_1^2}{G} \int^x dx x^{-2} \left\{ \left(\frac{\frac{\delta(x)}{16}}
{1 + \frac{\delta(x)}{4} x}\right) \right. \left. \left. + \left( \frac{1}{\delta(x)^2} \ln \left( 1 + \frac{\delta(x)}{4} x \right) 
 - \frac{1}{\delta(x)} \frac{\frac{1}{4} x }
{1 + \frac{\delta(x)}{4} x} \right) 
\delta' (x) \right\}\right|_{
x=H_1^2\left( \frac{\dot{L}_\mathrm{p}}{L_\mathrm{p}} - \frac{1}{L_\mathrm{p}} \right)^{-2}} \nonumber\\
=& \frac{4 H_1^2}{G} \int^x dx x^{-2} \left\{ \left(\frac{\frac{\delta(x)}{16}}
{1 + \frac{\delta(x)}{4} x}\right) \right. \left. \left. + \left( \frac{1}{\delta(x)^2} \ln \left( 1 + \frac{\delta(x)}{4} x \right) 
 - \frac{1}{\delta(x)} \frac{\frac{1}{4} x }
{1 + \frac{\delta(x)}{4} x} \right) 
\delta' (x) \right\}\right|_{
x=H_1^2\left( \frac{\dot{L}_\mathrm{f}}{L_\mathrm{f}} + \frac{1}{L_\mathrm{f}} \right)^{-2}} \, ,
\end{align}
and
\begin{align}
\frac{3c^2}{\kappa^2 L^2_\mathrm{SM}} =& \left. \frac{3}{8\pi G}
\left\{\left( \frac{\dot{L}_\mathrm{p}}{L_\mathrm{p}} - \frac{1}{L_\mathrm{p}} \right)^{2} + H_1^2 f(x)\right\}
\right|_{x=H_1^2\left( \frac{\dot{L}_\mathrm{p}}{L_\mathrm{p}} - \frac{1}{L_\mathrm{p}} \right)^{-2}}\nonumber\\
=& \left. \frac{3}{8\pi G}
\left\{\left( \frac{\dot{L}_\mathrm{f}}{L_\mathrm{f}} + \frac{1}{L_\mathrm{f}} \right)^{2} + H_1^2 f(x)\right\}
\right|_{x=H_1^2\left( \frac{\dot{L}_\mathrm{f}}{L_\mathrm{f}} + \frac{1}{L_\mathrm{f}} \right)^{-2}} \, ,
\label{Sm-extended3}
\end{align}
respectively, with $f(x)$ is shown in Eq.~(\ref{f}). Therefore even for the extended case, 
the holographic energies coming from the Tsallis entropy, the R\'{e}nyi entropy and the Sharma-Mittal entropy 
can be expressed by the general infrared cutoff
in \cite{Nojiri:2005pu}. Interestingly, the corresponding cut-offs are determined 
in terms of the particle horizon and its derivative or in terms of the future horizon and its derivative.

\section{Some other DE models and their equivalence with generalized HDE}\label{sec_V}
Besides the Tsallis, R\'{e}nyi and Sharma-Mittal entropic DE scenario, some other DE models, in particular the Quintessence 
\cite{Halliwell:1986ja,Barreiro:1999zs,Rubano:2001su,
Sangwan:2018zpz} and the Ricci-DE models \cite{Gao:2007ep,Zhang:2009un,delCampo:2013hka}, will 
take part in the present analysis. 
Their equivalence with the generalized HDE and the corresponding cut-offs are discussed in the following two subsections respectively.

\subsection{Quintessence dark energy}

The present observation indicates that the equation of state parameter at the present universe is close to $\omega \simeq -1$, however, it says a little 
about the time evolution of $\omega$, and thus we can broaden our situation and consider a dark energy model where the equation of state changes with 
time. Such kind of dark energy models are the scalar field dark energy models where the dynamics of the scalar field over the FLRW 
space-time leads to a evolving EoS of the universe. So far, a wide amount of scalar field dark energy models have been proposed, 
these include Quintessence, Phantoms, K-essence, Tachyon, Dilatonic dark energy etc.

In this section, we consider the Quintessence dark energy (QDE) model and aim to show that QDE is equivalent to the generalized 
holographic dark energy model where 
$L_\mathrm{IR} = L_\mathrm{IR} \left( L_\mathrm{p},\dot{L}_\mathrm{p},\ddot{L}_\mathrm{p},L_\mathrm{f},\dot{L}_\mathrm{f},\ddot{L}_\mathrm{f} \right)$. The QDE action is given by, 
\begin{eqnarray}
 S = \int d^4x \sqrt{-g}\left[\frac{R}{16\pi G} - \frac{1}{2}g^{\mu\nu}\partial_{\mu}\phi\partial_{\nu}\phi - V(\phi)\right]\, ,
 \label{Q-action}
\end{eqnarray}
where $\phi$ is the Quintessence scalar field and $V(\phi)$ is its potential. The presence of the potential 
is important in the dark energy context, otherwise the energy density and pressure of the scalar field become equal, which 
in turn leads to a decelerating expansion of the universe. In particular, with $V(\phi) = 0$, the FLRW 
scale factor of the universe evolves as $a(t) \sim t^{2/3}$ and thus the scalar field model without potential is not compatible 
with dark energy observations. The Quintessence potential has the following form \cite{Halliwell:1986ja,Barreiro:1999zs,Rubano:2001su,
Sangwan:2018zpz}, 
\begin{equation}
V(\phi) = V_0\exp{\left[-\sqrt{\frac{16\pi G}{p}} \phi\right]} \, ,
\label{Q-potential}
\end{equation}
with $V_0$ and $p$ are constants. The Quintessence model with the above exponential potential has been extensively studied 
in \cite{Sangwan:2018zpz} where it was shown that the potential of Eq.(\ref{Q-potential}) leads to a viable dark energy model in respect to 
SNIa, BAO and H(z) observations. However the most stringent constraints on the dark energy EoS parameter ($\omega_\mathrm{Q}$) 
comes from the BAO observations, in particular $-1 < \omega_\mathrm{Q} < -0.85$ \cite{Sangwan:2018zpz}. 

The FLRW equations correspond to the action (\ref{Q-action}) are,
\begin{align}
H^2=&\frac{8\pi G}{3}\left(\frac{1}{2}\dot{\phi}^2 + V(\phi)\right) \, , \nonumber\\
\dot{H}=&-4\pi G\dot{\phi}^2 \, , 
\label{Q-FRW equations}
\end{align}
where, due to the homogeneity, the scalar field is considered to be the function of time only. The first FLRW equation immediately leads to the 
Quintessence energy density as,
\begin{eqnarray}
 \rho_\mathrm{Q} = \frac{1}{2}\dot{\phi}^2 + V(\phi) = -\frac{\dot{H}}{8\pi G} + V(\phi)\label{Q-energy density}\, ,
\end{eqnarray}
where in the second line, we use $\dot{H} = -4\pi G\dot{\phi}^2$. Eq.(\ref{Q-energy density}) evidents that 
the Quintessence energy density is not symmetric with respect to the Hubble parameter, unlike to the case of entropic dark energy models 
(that we considered earlier) where the entropic energy density proves to be symmetric in respect to the Hubble parameter. The 
exponential form of the Quintessence potential (see Eq.~(\ref{Q-potential})) 
allows the following solutions of the Hubble parameter and the scalar field as,
\begin{eqnarray}
H = \frac{p}{t} \quad \mbox{and} \quad \phi(t) = 2\sqrt{\frac{p}{16\pi G}}\ln{\left(\frac{t}{t_0}\right)} \, ,
\label{Q-solutions}
\end{eqnarray}
respectively. Here $t_0$ being a fudicial time and $V_0$, $p$, and $t_0$ are related by the following constraint equation,
\begin{eqnarray}
3p - 1 = V_0t_0^2\left(\frac{8\pi G}{p}\right)\, .
\label{Q-conditions}
\end{eqnarray}
Furthermore the evolution of the Hubble parameter clearly indicates 
that in order to get an accelerating expansion of the universe, the parameter $p$ is constrained to be $p > 1$. By using 
Eqs.~(\ref{Q-potential}) and (\ref{Q-solutions}), we can express the Quintessence potential in terms of the Hubble parameter as follows,
\begin{eqnarray}
V(\phi) = \left(3 - \frac{1}{p}\right)\frac{H^2}{8\pi G}\, .
\label{Q-potential2}
\end{eqnarray}
Plugging back the above expression into Eq.~(\ref{Q-energy density}), we get $\rho_\mathrm{Q}$ in terms of $H$ and $\dot{H}$ as,
\begin{eqnarray}
\rho_\mathrm{Q} = \frac{1}{8\pi G}\left\{\left(3 - \frac{1}{p}\right)H^2 - \dot{H}\right\}\, .
\label{Q-dark energy2}
\end{eqnarray}
Furthermore the pressure in the present context is given by,
\begin{eqnarray}
 p_\mathrm{Q} = -\frac{\dot{H}}{8\pi G} - V(\phi) = -\frac{1}{8\pi G}\left\{\left(3 - \frac{1}{p}\right)H^2 + \dot{H}\right\}\, ,
 \label{pQ}
\end{eqnarray}
which, along with Eq.(\ref{Q-dark energy2}) immediately leads to the corresponding EoS parameter as,
\begin{eqnarray}
 \omega_\mathrm{Q} = -\frac{\left\{\left(3 - \frac{1}{p}\right)H^2 + \dot{H}\right\}}{\left\{\left(3 - \frac{1}{p}\right)H^2 - \dot{H}\right\}}~~.
 \label{eosQ}
\end{eqnarray}
Having set the stage, now we are in a position to show the equivalence between QDE and generalized holographic dark energy model. The comparison 
of Eqs.~(\ref{Q-dark energy2}) and (\ref{basic}) immediately lead to the equivalent holographic cut-off ($L_\mathrm{Q}$) corresponds to the QDE as follows,
\begin{align}
\frac{3c^2}{\kappa^2 L^2_\mathrm{Q}} =& 
\frac{1}{8\pi G}\left\{\left(3 - \frac{1}{p}\right)\left( \frac{\dot{L}_\mathrm{p}}{L_\mathrm{p}} - \frac{1}{L_\mathrm{p}} \right)^2 
 - \left( \frac{\ddot{L}_\mathrm{p}}{L_\mathrm{p}} - \frac{\dot{L}_\mathrm{p}^2}{L_\mathrm{p}^2} + \frac{\dot{L}_\mathrm{p}}{L_\mathrm{p}^2} \right)\right\}\nonumber\\
=& \frac{1}{8\pi G}\left\{\left(3 - \frac{1}{p}\right)\left( \frac{\dot{L}_\mathrm{f}}{L_\mathrm{f}} + \frac{1}{L_\mathrm{f}} \right)^2 
 - \left( \frac{\ddot{L}_\mathrm{f}}{L_\mathrm{f}} - \frac{\dot{L}_\mathrm{f}^2}{L_\mathrm{f}^2} 
 - \frac{\dot{L}_\mathrm{f}}{L_\mathrm{f}^2} \right)\right\}\, .
\label{Q-GHDE}
\end{align}
Thereby the QDE can be equivalently mapped to the generalized holographic dark energy model where the cut-off is the function of 
$L_\mathrm{p}$, $\dot{L}_\mathrm{p}$, $\ddot{L}_\mathrm{p}$ or the function of $L_\mathrm{f}$, $\dot{L}_\mathrm{f}$, $\ddot{L}_\mathrm{f}$. 
Furthermore, the EoS parameter ($\Omega_\mathrm{hol}^{(Q)}$) corresponds to the hologrphic cut-off $L_\mathrm{Q}$ is given by,
\begin{eqnarray}
 \Omega_\mathrm{hol}^{(Q)} = -1 - \left(\frac{2}{3HL_\mathrm{Q}}\right)\frac{dL_\mathrm{Q}}{dt}~,
 \label{eos-holQ}
\end{eqnarray}
where $L_\mathrm{Q}$ is shown above. Clearly, in accordance of Eq.(\ref{HLL}), $\Omega_\mathrm{hol}^{(Q)}$ becomes equivalent 
to the $\omega_\mathrm{Q}$ of Eq.(\ref{eosQ}). Such equivalence leads to the fact that similar to the Quintessence energy density, the 
holographic energy density coming from the cut-off $L_\mathrm{Q}$ also provides a good dark energy model of our universe.\\

\subsection{Ricci dark energy}
In this section, we intend to establish that the Ricci dark energy (RDE) model has a direct equivalence to the generalized holographic dark energy model. 
The RDE model \cite{Gao:2007ep,Zhang:2009un,delCampo:2013hka} catches a special attention as the dark energy density in this context 
has a geometric origin, in particular the dark energy density is given by,
\begin{eqnarray}
\rho_\mathrm{RD} = \frac{\alpha}{16\pi}R = \frac{3\alpha}{8\pi}\left(\dot{H} + 2H^2\right)\, ,\label{R-dark energy1}
\end{eqnarray}
with $R$ being the space-time Ricci scalar and $\alpha$ is a model parameter. The above expression of $\rho_\mathrm{RD}$ along with 
its conservation equation lead to the corresponding EoS parameter as,
\begin{eqnarray}
 \omega_\mathrm{RD} = -1 + \frac{\alpha\left(1+z\right)}{8\pi}\frac{d}{dz}\left[\ln{\left(\dot{H} + 2H^2\right)}\right]~,
 \label{eosRD}
\end{eqnarray}
where we use the explicit form of $\rho_\mathrm{RD}$ and $z = a^{-1} - 1$ is known as the red-shift factor. 
It is evident that the parameter $\alpha$ actually controls the evolution of the $\omega_\mathrm{RD}$ and hence the universe's evolution. 
In particular, 
it has been showed in \cite{Zhang:2009un} that for $1/2 < \alpha < 1$, 
the RDE has EoS $-1 < \omega_\mathrm{RD} < -1/3$ and for the case $\alpha < 1/2$, the RDE start from quintessence-like and evolves to phantom-like. 
In regard to the observational compatibility of RDE, the parameter 
$\alpha$ is constrained by $\alpha = 0.394^{+0.152}_{-0.106}$ from SNIa only ($1\sigma$), however a 
joint analysis of the SNIa+CMB+BAO observations gives a much tighter constraint on $\alpha$ as $\alpha = 0.359^{+0.024}_{-0.025}$ \cite{Zhang:2009un}.

Eqs.~(\ref{R-dark energy1}) and (\ref{basic}) indicate that the RDE has a direct equivalence to the generalized holographic dark energy model, 
where the corresponding the cut-off ($L_\mathrm{RD}$) can be expressed as,
\begin{eqnarray}
\frac{3c^2}{\kappa^2 L^2_\mathrm{RD}} 
= \frac{3\alpha}{8\pi}\left\{\frac{\ddot{L}_\mathrm{p}}{L_\mathrm{p}} + \frac{\dot{L}_\mathrm{p}^2}{L_\mathrm{p}^2} - 3\frac{\dot{L}_\mathrm{p}}{L_\mathrm{p}^2} 
+ \frac{2}{L_\mathrm{p}^2}\right\} \, ,
\label{R1-GHDE}
\end{eqnarray}
in terms of $L_\mathrm{p}$, $\dot{L}_\mathrm{p}$ and $\ddot{L}_\mathrm{p}$. Similarly, the $L_\mathrm{RD}$ in terms of future horizon and its derivatives is given by,
\begin{eqnarray}
\frac{3c^2}{\kappa^2 L^2_\mathrm{RD}} 
= \frac{3\alpha}{8\pi}\left\{\frac{\ddot{L}_\mathrm{f}}{L_\mathrm{f}} + \frac{\dot{L}_\mathrm{f}^2}{L_\mathrm{f}^2} + 3\frac{\dot{L}_\mathrm{f}}{L_\mathrm{f}^2} 
+ \frac{2}{L_\mathrm{f}^2}\right\}\, .
\label{R2-GHDE}
\end{eqnarray}
Such holographic cut-offs establish a symmetry between the RDE and generalized HDE. 
Furthermore a modified form of RDE has been proposed in \cite{Granda:2008dk}, where the dark energy density comes with the following form,
\begin{eqnarray}
\rho_\mathrm{RD} = 3\left(\alpha H^2 + \beta \dot{H}\right) \, ,
\label{R-dark energy2}
\end{eqnarray}
with $\alpha$ and $\beta$ are two parameters. The comparison of the above equation with Eq.~(\ref{basic}) immediately 
leads to the equivalence holographic cut-off (in terms of $L_\mathrm{p}$ and its derivatives or in terms of $L_\mathrm{f}$ and its derivatives) 
corresponds to the modified RDE as,
\begin{align}
\frac{3c^2}{\kappa^2 L^2_\mathrm{RD}} 
=&3\left\{\alpha\left( \frac{\dot{L}_\mathrm{p}}{L_\mathrm{p}} - \frac{1}{L_\mathrm{p}} \right)^2 
+ \beta\left( \frac{\ddot{L}_\mathrm{p}}{L_\mathrm{p}} - \frac{\dot{L}_\mathrm{p}^2}{L_\mathrm{p}^2} + \frac{\dot{L}_\mathrm{p}}{L_\mathrm{p}^2} \right)\right\}\nonumber\\
=&3\left\{\alpha\left( \frac{\dot{L}_\mathrm{f}}{L_\mathrm{f}} + \frac{1}{L_\mathrm{f}} \right)^2 
 + \beta\left( \frac{\ddot{L}_\mathrm{f}}{L_\mathrm{f}} - \frac{\dot{L}_\mathrm{f}^2}{L_\mathrm{f}^2} 
 - \frac{\dot{L}_\mathrm{f}}{L_\mathrm{f}^2} \right)\right\}\, ,
 \label{R3-GHDE}
\end{align}
where we use Eq.~(\ref{HLL}). The EoS parameter ($\Omega_\mathrm{hol}^{(RD)}$) corresponds to the holographic energy density 
$\rho_\mathrm{hol} = 3c^2/\left(\kappa^2 L^2_\mathrm{RD}\right)$ arises from its conservation relation, in particular we get
\begin{eqnarray}
 \Omega_\mathrm{hol}^{(RD)} = -1 - \left(\frac{2}{3HL_\mathrm{RD}}\right)\frac{dL_\mathrm{RD}}{dt}~,
 \label{eos-holRD}
\end{eqnarray}
which, due to Eq.(\ref{HLL}), evidents to be equivalent with the $\omega_\mathrm{RD}$ of Eq.(\ref{eosRD}). Thus similar to the RDE, 
the holographic energy density having the cut-off $L_\mathrm{RD}$ proves to be a viable dark energy model of the universe in regard to 
the SNIa+CMB+BAO observations. 
Therefore the RDE and the modified RDE may be regarded as certain candidates of the generalized holographic dark energy family, with 
the respective holographic cut-offs given by Eqs.~(\ref{R1-GHDE}), (\ref{R2-GHDE}), and (\ref{R3-GHDE}), respectively.\\

Before concluding we consider the scale invariant cosmological field equations \cite{Maeder:2021rqd} 
and investigate its holographic correspondence. The 
said field equations are given by \cite{Maeder:2021rqd},
\begin{eqnarray}
 H^2&=&\frac{8\pi G}{3}\rho - 2H\frac{\dot{\lambda}}{\lambda}~,\label{SI-1}\\
 2\dot{H} + 3H^2&=&-8\pi Gp - 4H\frac{\dot{\lambda}}{\lambda}~,
 \label{SI-2}
\end{eqnarray}
where $\rho$ and $p$ represent the energy density and pressure of the matter components. Moreover $\lambda$ parametrizes the scale invariance, which 
varies with the expansion of the universe, i.e $\lambda = \lambda(a(t))$. Here it may be mentioned that the authors of \cite{Maeder:2021rqd} proposed 
an inflationary scenario in the context of scale invariance cosmology, 
in which case the matter components are provided by a slow rolling scalar field, in particular 
$\rho = \frac{1}{2}C\left(\dot{\Psi}^2 + U(\Psi)\right)$ and $p = \frac{1}{2}C\left(\dot{\Psi}^2 - U(\Psi)\right)$, where 
$\Psi$ is a scalar field, $U(\Psi)$ being its potential and $C$ is a constant (for more information about $C$, see \cite{Maeder:2021rqd}). 
Clearly the field Eqs.(\ref{SI-1}) and (\ref{SI-2}) can be equivalently mapped to the holographic cosmological scenario, where the holographic cut-off 
and the corresponding EoS parameter are given by,
\begin{eqnarray}
 \frac{3c^2}{\kappa^2L_\mathrm{SI}^2}&=&\rho(a) - \frac{3}{4\pi G}\left(\frac{\dot{\lambda}(a)}{\lambda(a)}\right)
 \left(\frac{\dot{L}_\mathrm{p}}{L_\mathrm{p}} - \frac{1}{L_\mathrm{p}}\right)~,\label{SI-3}\\
 \mathrm{or}~,~\frac{3c^2}{\kappa^2L_\mathrm{SI}^2}&=&\rho(a) - \frac{3}{4\pi G}\left(\frac{\dot{\lambda}(a)}{\lambda(a)}\right)
 \left(\frac{\dot{L}_\mathrm{f}}{L_\mathrm{f}} + \frac{1}{L_\mathrm{f}}\right)~,
 \label{SI-4}
\end{eqnarray}
and 
\begin{eqnarray}
 \Omega_\mathrm{hol}^{(SI)} = -1 - \left(\frac{2}{3HL_\mathrm{SI}}\right)\frac{dL_\mathrm{SI}}{dt}~.
 \label{eos-holSI}
\end{eqnarray}
Eqs.(\ref{SI-3}) and (\ref{SI-4}) represent the $L_\mathrm{SI}$ in terms of $L_\mathrm{p}$ (and its derivative) and 
$L_\mathrm{f}$ (and its derivative) respectively. The EoS parameter in Eq.(\ref{eos-holSI}) satisfies the conservation relation 
of the holographic energy density $\rho_\mathrm{hol} = 3c^2/\left(\kappa^2 L^2_\mathrm{SI}\right)$. As a whole, 
the scale invariant cosmological model (described by Eqs.(\ref{SI-1}) and (\ref{SI-2})) has a holographic correspondence, with the cut-off being 
given in Eq.(\ref{eos-holSI}).

At this stage it deserves mentioning that as far as we could see on many examples of modified gravity, scalar-tensor theory or gravity with 
fluids the corresponding FLRW equations can be always mapped to holographic cosmology with specific 
IR cut-off. However, of course the physical nature of such cut-off remains to be obscure.

\section{Conclusion}

Dark energy (DE) is one of the most puzzled issues in modern cosmology. In particular, DE may even be an issue of quantum gravity. In this regard, 
the holographic principle, one of the important cornerstones of quantum gravity, plays an important role in describing the dark energy of 
our universe. Based on the holographic principle and on the dimensional analysis, the theory of holographic dark energy (HDE) 
has been formulated, where the dark energy density is proportional to the inverse squared if the infrared cut-off. 
The holographic cut-off is usually considered to be same as 
the particle horizon or the future horizon. It may be stressed that instead of adding a term into the Lagrangian, the HDE 
is based on the holographic principle and on the dimensional analysis and this makes the HDE significantly different than the other theory of DE. 
In \cite{Nojiri:2005pu}, a generalized HDE has been proposed where the cut-off ($L_\mathrm{IR}$) is generalized to be a function of 
particle horizon ($L_\mathrm{p}$) and its derivatives of any order or a function of future horizon ($L_\mathrm{f}$) and its derivatives of any order, 
in particular $L_\mathrm{IR} = L_\mathrm{IR} \left(L_\mathrm{p}, \dot L_\mathrm{p}, 
\ddot L_\mathrm{p}, \cdots, L_\mathrm{f}, \dot L_\mathrm{f}, \cdots, a\right)$. Evidently, with such 
generalized form of the $L_\mathrm{IR}$, the phenomenology of the generalized HDE becomes more richer.

Based on the formalism of the generalized HDE, we showed that a wide class of dark energy models can be regarded as different candidates of the 
generalized holographic dark energy family with respective cut-offs. In this regard, we first considered several entropic 
DE models, in particular the Tsallis entropic DE, the R\'{e}nyi entropic DE, and the Sharma-Mittal entropic DE, and showed that they 
are indeed equivalent to the generalized HDE model, where the corresponding cut-offs are determined in terms of $L_\mathrm{p}$, $\dot{L}_\mathrm{p}$ or in terms of 
$L_\mathrm{f}$, $\dot{L}_\mathrm{f}$, respectively. Such equivalence between the entropic DE and the generalized HDE are established for two cases: 
(1) in the first case, the exponents of the respective entropy functions are regarded to be constant, while in the second case 
(2) the exponents vary with cosmic time, particularly the exponents are considered to depend on the evolving Hubble parameter. 
Here it may be mentioned that for the entropic DE models, the equivalent holographic cut-offs depend up-to the $first$ derivative 
of $L_\mathrm{p}$ or $L_\mathrm{f}$. Besides such entropic DE models, some other DE models like - (1) the Quintessence model where a minimally 
coupled scalar field with an exponential potential serves the dark energy density and (2) the Ricci DE where the space-time 
curvature provides the dark energy density - are also proved to be equivalent with the generalized HDE. The equivalent holographic cut-off 
corresponds to the Quintessence as well as to the Ricci DE model depends on either $L_\mathrm{p}$, $\dot{L}_\mathrm{p}$, $\ddot{L}_\mathrm{p}$ or $L_\mathrm{f}$, $\dot{L}_\mathrm{f}$, $\ddot{L}_\mathrm{f}$. 
It may be noted that for both the Quintessence and Ricci DE models, the equivalent cut-offs depend up-to the $second$ 
derivative of $L_\mathrm{p}$ or $L_\mathrm{f}$, unlike to that of the entropic DE models where, as mentioned earlier, 
the corresponding $L_\mathrm{IR}$ depends at most on the $first$ derivative of the $L_\mathrm{p}$ or $L_\mathrm{f}$ respectively. 
Finally it deserves mentioning that in all the cases, we determine the effective EoS parameter for the DE models and the 
corresponding generalized HDE models, where the EoS parameter are represented by $\omega_\mathrm{i}$ and $\Omega_\mathrm{hol}^{(i)}$ respectively 
(with $i$ denotes the various cases we considered). As a result, we found that $\omega_\mathrm{i} \equiv \Omega_\mathrm{hol}^{(i)}$, which 
further confirms the equivalence between various DE models and the respective generalized HDE models. 
This indicates a symmetry between the generalized HDE and different DE models.

In summary, a wide class of dark energy models including the entropic DE models are found to be equivalent with the $generalized$ HDE, with the 
corresponding cut-offs being determined in terms of the particle horizon and its derivatives or in terms of the 
future horizon and its derivatives. However, the understanding for the choice of fundamental viable cut-off still remains to be a debatable topic. 
The comparison of such cut-offs for realistic description of the universe evolution may help in 
better understanding of holographic principle. Furthermore, it is interesting to note that recently holographic 
inflation \cite{Nojiri:2019kkp} was proposed with above generalized holographic cut-off. 
Our considerations indicate that similar equivalence may be established 
between different inflationary theories and holographic inflationary model with generalized cut-off.

\appendix
\section{Detailed derivations of Sec.[\ref{Sec4}]}\label{appendix}

\subsection{Derivation of Eq.(\ref{Tslls17})}
The Tsallis entropy with varying exponent, in particular $\delta = \delta(x)$ where $x = H_1^2/H^2$, is given by,
\begin{eqnarray}
 S_T = \frac{A_0}{4G}\left(\frac{A}{A_0}\right)^{\delta(x)}\label{app1-1}~~.
\end{eqnarray}
Therefore,
\begin{eqnarray}
 \frac{dS_T}{dt}&=&\frac{\partial S}{\partial A}\frac{dA}{dt} + \frac{\partial S}{\partial x}\frac{dx}{dt}\nonumber\\
 &=&-\frac{1}{4G}\left(\frac{8\pi\dot{H}}{H^3}\right)\left(\frac{A}{A_0}\right)^{\delta(x)-1}
 \left\{\delta(x) + \frac{H_1^2}{H^2}\ln{\left(\frac{H_1^2}{H^2}\right)}\delta'(x)\right\}\nonumber\\
 &=&-\frac{1}{4G}\left(\frac{8\pi}{H^3}\right)\left(\frac{H_1^2}{H^2}\right)^{\delta(x)-1}
 \left\{\delta(x) + \frac{H_1^2}{H^2}\ln{\left(\frac{H_1^2}{H^2}\right)}\delta'(x)\right\}\dot{H}~~,\nonumber\\
 \label{app1-2}
\end{eqnarray}
where in the second equality of the above equation, we use $A = 4\pi r_h^2$ and $r_h = H^{-1}$. Using the 
above expression of $\frac{dS}{dt}$ into the first law of thermodynamics in Eq.(\ref{Tslls2}), one gets
\begin{eqnarray}
 \left\{\delta(x) + \frac{H_1^2}{H^2}\ln{\left(\frac{H_1^2}{H^2}\right)}\delta'(x)\right\}\left(\frac{H_1^2}{H^2}\right)^{\delta(x)-1}\dot{H} 
 = -4\pi G\left(\rho + p\right)~~,
 \label{app1-3}
\end{eqnarray}
with $\rho$ and $p$ represent the energy density and pressure of the matter contents, and obey the conservation relation: 
$\dot{\rho} + 3H\left(\rho + p\right) = 0$. In accordance of this conservation relation, we can integrate Eq.(\ref{app1-3}) as,
\begin{eqnarray}
 \int dx~x^{\delta(x)-1}\left\{\delta(x) + x\delta'(x)\ln{x}\right\}H\frac{dH}{dx} = \frac{4\pi G}{3}\rho + \frac{\Lambda}{6}~~,
 \label{app1-4}
\end{eqnarray}
where $\Lambda$ is the constant of integration. The expression $x = H_1^2/H^2$ immediately yields
\begin{eqnarray}
 H\frac{dH}{dx} = -\frac{H_1^2}{2x^2}~,\label{app1-5}
\end{eqnarray}
by plugging which into Eq.(\ref{app1-4}), we obtain,
\begin{eqnarray}
 -H_1^2\int dx \left\{\delta(x) + x\delta'(x)\ln{x}\right\}x^{\delta(x)-3}&=&\frac{8\pi G}{3}\rho + \frac{\Lambda}{3}\nonumber\\
 \Rightarrow -H_1^2\left\{x^{\delta(x)-2} + 2\int dx~x^{\delta(x)-3}\right\}\bigg|_{x=H_1^2/H^2}&=&\frac{8\pi G}{3}\rho + \frac{\Lambda}{3}~~,
 \label{app1-6}
\end{eqnarray}
which is written in Eq.(\ref{Tslls17}).

\subsection{Derivation of Eq.(\ref{RFRW0001})}
The R\'{e}nyi entropy with varying $\delta = \delta(x)$ is given by,
\begin{eqnarray}
 S_\mathrm{R} = \frac{A_0}{G\delta(x)}\ln{\left(1 + \frac{\delta(x)}{4}\left(\frac{A}{A_0}\right)\right)}~~.
 \label{app2-1}
\end{eqnarray}
Therefore,
\begin{eqnarray}
 \frac{dS_\mathrm{R}}{dt}&=&\frac{\partial S_{R}}{\partial A}\frac{dA}{dt} + \frac{\partial S_{R}}{\partial x}\frac{dx}{dt}\nonumber\\
 &=&\frac{1}{4G}\frac{1}{\left(1 + \frac{\delta(x)}{4}\left(\frac{A}{A_0}\right)\right)}\frac{dA}{dt} 
 + \frac{A_0}{G}\left\{-\frac{\delta'(x)}{\delta^2(x)}\ln{\left(1 + \frac{\delta(x)}{4}\left(\frac{A}{A_0}\right)\right)} 
 + \frac{\delta'(x)}{4\delta(x)\left(1 + \frac{\delta(x)}{4}\left(\frac{A}{A_0}\right)\right)}\left(\frac{A}{A_0}\right)\right\}\frac{dx}{dt}\nonumber\\
 &=&\frac{8\pi}{H^3}\left[-\frac{1}{4\left(1 + \frac{\delta(x)}{4}\left(\frac{A}{A_0}\right)\right)} 
 + \delta'(x)\left\{\frac{1}{\delta^2(x)}\ln{\left(1 + \frac{\delta(x)}{4}\left(\frac{A}{A_0}\right)\right)} 
 - \frac{x}{4\delta(x)\left(1 + \frac{\delta(x)}{4}\left(\frac{A}{A_0}\right)\right)}\right\}\right]\left(\frac{\dot{H}}{G}\right)~.
 \nonumber
\end{eqnarray}
Plugging the above expression into Eq.(\ref{Tslls2}), one gets the evolution of $\dot{H}$ in the present case as,
\begin{eqnarray}
 \left[\frac{1}{\left(1 + \frac{x\delta(x)}{4}\right)} - 4\delta'(x)
 \left\{\frac{1}{\delta^2(x)}\ln{\left(1 + \frac{x\delta(x)}{4}\right)} - \frac{x}{4\delta(x)\left(1 + \frac{x\delta(x)}{4}\right)}\right\}\right]\dot{H} 
 = -4\pi G\left(\rho + p\right)~.
 \label{app2-2}
\end{eqnarray}
Using the conservation relation $\dot{\rho} + 3H\left(\rho + p\right) = 0$, we integrate Eq.(\ref{app2-2}) to yield,
\begin{eqnarray}
 \int dx\left[\frac{1}{\left(1 + \frac{x\delta(x)}{4}\right)} - 4\delta'(x)
 \left\{\frac{1}{\delta^2(x)}\ln{\left(1 + \frac{x\delta(x)}{4}\right)} - \frac{x}{4\delta(x)\left(1 + \frac{x\delta(x)}{4}\right)}\right\}\right]
 H\frac{dH}{dx} = \frac{4\pi G}{3}\rho + \frac{\Lambda}{6}~,
 \label{app2-3}
\end{eqnarray}
where $\Lambda$ represents the constant of integration. Due to the expression of $H\frac{dH}{dx}$ shown in Eq.(\ref{app1-5}), the above equation takes the 
following form,
\begin{eqnarray}
 H^2 = \frac{8\pi G}{3} \rho + \frac{\Lambda}{3} 
+ 4H_1^2 \left. \int^x dx x^{-2} \left\{ \left(\frac{\frac{\delta(x)}{16}}
{1 + \frac{\delta(x)}{4} x}\right)
+ \left( \frac{1}{\delta(x)^2} \ln \left( 1 + \frac{\delta(x)}{4} x \right) 
 - \frac{1}{\delta(x)} \frac{\frac{1}{4} x }
{1 + \frac{\delta(x)}{4} x} \right) 
\delta' (x) \right\}\right|_{x=\frac{H_1^2}{H^2}}~,\nonumber\\
\label{app2-4}
\end{eqnarray}
which is written in Eq.(\ref{RFRW0001}).

\subsection{Derivation of Eq.(\ref{SM-extended1})}
The Sharma-Mittal entropy with varying exponent, in particular with $\alpha = \alpha(x)$ and $\delta = \mathrm{constant}$, is given by
\begin{eqnarray}
 S_\mathrm{SM} = \frac{A_0}{G\alpha(x)}\left\{\left( 
1 + \frac{\delta}{4}\left(\frac{A}{A_0} \right)\right)^{\frac{\alpha(x)}{\delta}} - 1\right\}~.
\label{app3-1}
\end{eqnarray}
Thereby,
\begin{eqnarray}
 \frac{dS_\mathrm{SM}}{dt}&=&\frac{\partial S_{SM}}{\partial A}\frac{dA}{dt} + \frac{\partial S_{SM}}{\partial x}\frac{dx}{dt}\nonumber\\
 &=&\frac{1}{4G}\left(1 + \frac{\delta}{4}\left(\frac{A}{A_0}\right)\right)^{\frac{\alpha(x)}{\delta}-1}\frac{dA}{dt}\nonumber\\ 
 &+&\frac{A_0}{G}\left[-\frac{\alpha'(x)}{\alpha^2(x)}\left\{\left(1 + \frac{\delta}{4}\left(\frac{A}{A_0}\right)\right)^{\frac{\alpha(x)}{\delta}} 
 - 1\right\} + \frac{\alpha'(x)}{\alpha(x)\delta}\left(1 + \frac{\delta}{4}\left(\frac{A}{A_0}\right)\right)^{\frac{\alpha(x)}{\delta}}
 \ln{\left(1 + \frac{\delta}{4}\left(\frac{A}{A_0}\right)\right)}\right]\frac{dx}{dt}\nonumber\\
 &=&\frac{8\pi}{H^3}\bigg[-\frac{1}{4}\left(1 + \frac{\delta}{4}\left(\frac{A}{A_0}\right)\right)^{\frac{\alpha(x)}{\delta}-1}\nonumber\\ 
 &+&\frac{\alpha'(x)}{\alpha^2(x)}\left\{\left(1 + \frac{\delta}{4}\left(\frac{A}{A_0}\right)\right)^{\frac{\alpha(x)}{\delta}} - 1\right\} 
 - \frac{\alpha'(x)}{\alpha(x)\delta}\left(1 + \frac{\delta}{4}\left(\frac{A}{A_0}\right)\right)^{\frac{\alpha(x)}{\delta}}
 \ln{\left(1 + \frac{\delta}{4}\left(\frac{A}{A_0}\right)\right)}\bigg]\left(\frac{\dot{H}}{G}\right)~.
 \label{app3-2}
\end{eqnarray}
The above expression of $\frac{dS_\mathrm{SM}}{dt}$ along with the first law of thermodynamics in Eq.(\ref{Tslls2}) lead to the following 
equation of $\dot{H}$,
\begin{eqnarray}
 \bigg[\left(1 + \frac{\delta x}{4}\right)^{\frac{\alpha(x)}{\delta}-1} - \frac{4\alpha'(x)}{\alpha^2(x)}
 \left\{\left(1 + \frac{\delta x}{4}\right)^{\frac{\alpha(x)}{\delta}} - 1\right\} 
 + \frac{4\alpha'(x)}{\alpha(x)\delta}\left(1 + \frac{\delta x}{4}\right)^{\frac{\alpha(x)}{\delta}}
 \ln{\left(1 + \frac{\delta x}{4}\right)}\bigg]\dot{H} = -4\pi G\left(\rho + p\right)~,\nonumber\\
 \label{app3-3}
\end{eqnarray}
on integrating which, we obtain Eq.(\ref{SM-extended1}). Here it may be mentioned that to derive Eq.(\ref{SM-extended1}), we use 
$H\frac{dH}{dx} = -H_1^2/(2x^2)$ and the conservation relation of $\rho$.

\begin{acknowledgments}
This work is supported by the JSPS Grant-in-Aid for Scientific Research (C)
No. 18K03615 (S.N.).
\end{acknowledgments}


\begin{thebibliography}{99}

\bibitem{tHooft:1993dmi}
G.~'t Hooft,
Conf.\ Proc.\ C {\bf 930308} (1993) 284
[gr-qc/9310026].

\bibitem{Susskind:1994vu}
L.~Susskind,
J.\ Math.\ Phys.\  {\bf 36} (1995) 6377
doi:10.1063/1.531249
[hep-th/9409089].

\bibitem{Witten:1998qj}
E.~Witten,
Adv.\ Theor.\ Math.\ Phys.\  {\bf 2} (1998) 253
doi:10.4310/ATMP.1998.v2.n2.a2
[hep-th/9802150].

\bibitem{Bousso:2002ju}
R.~Bousso,
Rev.\ Mod.\ Phys.\  {\bf 74} (2002) 825
doi:10.1103/RevModPhys.74.825
[hep-th/0203101].


\bibitem{Li:2004rb}
M.~Li,
Phys.\ Lett.\ B {\bf 603} (2004) 1
doi:10.1016/j.physletb.2004.10.014
[hep-th/0403127].

\bibitem{Li:2011sd}
M.~Li, X.~D.~Li, S.~Wang and Y.~Wang,
Commun. Theor. Phys. \textbf{56} (2011), 525-604
doi:10.1088/0253-6102/56/3/24
[arXiv:1103.5870 [astro-ph.CO]].

\bibitem{Wang:2016och}
S.~Wang, Y.~Wang and M.~Li,
Phys.\ Rept.\  {\bf 696} (2017) 1
doi:10.1016/j.physrep.2017.06.003
[arXiv:1612.00345 [astro-ph.CO]].

\bibitem{Pavon:2005yx}
D.~Pavon and W.~Zimdahl,
Phys.\ Lett.\ B {\bf 628} (2005) 206
doi:10.1016/j.physletb.2005.08.134
[gr-qc/0505020].

\bibitem{Nojiri:2005pu}
S.~Nojiri and S.~D.~Odintsov,
Gen.\ Rel.\ Grav.\  {\bf 38} (2006) 1285
doi:10.1007/s10714-006-0301-6
[hep-th/0506212].


\bibitem{Enqvist:2004xv}
K.~Enqvist and M.~S.~Sloth,
Phys.\ Rev.\ Lett.\  {\bf 93} (2004) 221302
doi:10.1103/PhysRevLett.93.221302
[hep-th/0406019].

\bibitem{Zhang:2005yz}
X.~Zhang,
Int.\ J.\ Mod.\ Phys.\ D {\bf 14} (2005) 1597
doi:10.1142/S0218271805007243
[astro-ph/0504586].

\bibitem{Guberina:2005fb}
B.~Guberina, R.~Horvat and H.~Stefancic,
JCAP {\bf 0505} (2005) 001
doi:10.1088/1475-7516/2005/05/001
[astro-ph/0503495].

\bibitem{Elizalde:2005ju}
E.~Elizalde, S.~Nojiri, S.~D.~Odintsov and P.~Wang,
Phys.\ Rev.\ D {\bf 71} (2005) 103504
doi:10.1103/PhysRevD.71.103504
[hep-th/0502082].

\bibitem{Ito:2004qi}
M.~Ito,
Europhys.\ Lett.\  {\bf 71} (2005) 712
doi:10.1209/epl/i2005-10151-x
[hep-th/0405281].

\bibitem{Gong:2004cb}
Y.~g.~Gong, B.~Wang and Y.~Z.~Zhang,
Phys.\ Rev.\ D {\bf 72} (2005) 043510
doi:10.1103/PhysRevD.72.043510
[hep-th/0412218].

\bibitem{Saridakis:2007cy}
E.~N.~Saridakis,
Phys.\ Lett.\ B {\bf 660} (2008) 138
doi:10.1016/j.physletb.2008.01.004
[arXiv:0712.2228 [hep-th]].

\bibitem{Gong:2009dc}
Y.~Gong and T.~Li,
Phys.\ Lett.\ B {\bf 683} (2010) 241
doi:10.1016/j.physletb.2009.12.040
[arXiv:0907.0860 [hep-th]].

\bibitem{BouhmadiLopez:2011xi}
M.~Bouhmadi-Lopez, A.~Errahmani and T.~Ouali,
Phys.\ Rev.\ D {\bf 84} (2011) 083508
doi:10.1103/PhysRevD.84.083508
[arXiv:1104.1181 [astro-ph.CO]].

\bibitem{Malekjani:2012bw}
M.~Malekjani,
Astrophys.\ Space Sci.\  {\bf 347} (2013) 405
doi:10.1007/s10509-013-1522-2
[arXiv:1209.5512 [gr-qc]].

\bibitem{Khurshudyan:2014axa}
M.~Khurshudyan, J.~Sadeghi, R.~Myrzakulov, A.~Pasqua and H.~Farahani,
Adv.\ High Energy Phys.\  {\bf 2014} (2014) 878092
doi:10.1155/2014/878092
[arXiv:1404.2141 [gr-qc]].

\bibitem{Khurshudyan:2016gmb}
M.~Khurshudyan,
Astrophys. Space Sci. \textbf{361} (2016) no.12, 392
doi:10.1007/s10509-016-2981-z

\bibitem{Landim:2015hqa}
R.~C.~G.~Landim,
Int.\ J.\ Mod.\ Phys.\ D {\bf 25} (2016) no.04,  1650050
doi:10.1142/S0218271816500504
[arXiv:1508.07248 [hep-th]].

\bibitem{Gao:2007ep}
C.~Gao, F.~Wu, X.~Chen and Y.~G.~Shen,
Phys.\ Rev.\ D {\bf 79} (2009) 043511
doi:10.1103/PhysRevD.79.043511
[arXiv:0712.1394 [astro-ph]].

\bibitem{Li:2008zq}
M.~Li, C.~Lin and Y.~Wang,
JCAP {\bf 0805} (2008) 023
doi:10.1088/1475-7516/2008/05/023
[arXiv:0801.1407 [astro-ph]].

\bibitem{Anagnostopoulos:2020ctz}
F.~K.~Anagnostopoulos, S.~Basilakos and E.~N.~Saridakis,
[arXiv:2005.10302 [gr-qc]].


\bibitem{Zhang:2005hs}
X.~Zhang and F.~Q.~Wu,
Phys.\ Rev.\ D {\bf 72} (2005) 043524
doi:10.1103/PhysRevD.72.043524
[astro-ph/0506310].

\bibitem{Li:2009bn}
M.~Li, X.~D.~Li, S.~Wang and X.~Zhang,
JCAP {\bf 0906} (2009) 036
doi:10.1088/1475-7516/2009/06/036
[arXiv:0904.0928 [astro-ph.CO]].

\bibitem{Feng:2007wn}
C.~Feng, B.~Wang, Y.~Gong and R.~K.~Su,
JCAP {\bf 0709} (2007) 005
doi:10.1088/1475-7516/2007/09/005
[arXiv:0706.4033 [astro-ph]].

\bibitem{Zhang:2009un}
X.~Zhang,
Phys.\ Rev.\ D {\bf 79} (2009) 103509
doi:10.1103/PhysRevD.79.103509
[arXiv:0901.2262 [astro-ph.CO]].

\bibitem{Lu:2009iv}
J.~Lu, E.~N.~Saridakis, M.~R.~Setare and L.~Xu,
JCAP {\bf 1003} (2010) 031
doi:10.1088/1475-7516/2010/03/031
[arXiv:0912.0923 [astro-ph.CO]].

\bibitem{Micheletti:2009jy}
S.~M.~R.~Micheletti,
JCAP {\bf 1005} (2010) 009
doi:10.1088/1475-7516/2010/05/009
[arXiv:0912.3992 [gr-qc]].

\bibitem{Huang:2004wt}
Q.~G.~Huang and Y.~G.~Gong,
JCAP {\bf 0408} (2004) 006
doi:10.1088/1475-7516/2004/08/006
[astro-ph/0403590].

\bibitem{Mukherjee:2017oom}
P.~Mukherjee, A.~Mukherjee, H.~Jassal, A.~Dasgupta and N.~Banerjee,
Eur. Phys. J. Plus \textbf{134} (2019) no.4, 147
doi:10.1140/epjp/i2019-12504-7
[arXiv:1710.02417 [astro-ph.CO]].

\bibitem{Nojiri:2017opc}
S.~Nojiri and S.~Odintsov,
Eur. Phys. J. C \textbf{77} (2017) no.8, 528
doi:10.1140/epjc/s10052-017-5097-x
[arXiv:1703.06372 [hep-th]].

\bibitem{Sharif:2019seo}
M.~Sharif and S.~Saba,
Symmetry \textbf{11} (2019) no.1, 92
doi:10.3390/sym11010092

\bibitem{Jawad:2018juh}
A.~Jawad, K.~Bamba, M.~Younas, S.~Qummer and S.~Rani,
Symmetry \textbf{10} (2018) no.11, 635
doi:10.3390/sym10110635

\bibitem{Horvat:2011wr}
R.~Horvat,
Phys. Lett. B \textbf{699} (2011), 174-176
doi:10.1016/j.physletb.2011.04.004
[arXiv:1101.0721 [hep-ph]].

\bibitem{Nojiri:2019kkp}
S.~Nojiri, S.~D.~Odintsov and E.~N.~Saridakis,
Phys. Lett. B \textbf{797} (2019), 134829
doi:10.1016/j.physletb.2019.134829
[arXiv:1904.01345 [gr-qc]].

\bibitem{Paul:2019hys}
T.~Paul,
EPL \textbf{127} (2019) no.2, 20004
doi:10.1209/0295-5075/127/20004
[arXiv:1905.13033 [gr-qc]].

\bibitem{Bargach:2019pst}
A.~Bargach, F.~Bargach, A.~Errahmani and T.~Ouali,
Int. J. Mod. Phys. D \textbf{29} (2020) no.02, 2050010
doi:10.1142/S0218271820500108
[arXiv:1904.06282 [hep-th]].

\bibitem{Elizalde:2019jmh}
E.~Elizalde and A.~Timoshkin,
Eur. Phys. J. C \textbf{79} (2019) no.9, 732
doi:10.1140/epjc/s10052-019-7244-z
[arXiv:1908.08712 [gr-qc]].

\bibitem{Oliveros:2019rnq}
A.~Oliveros and M.~A.~Acero,
EPL \textbf{128} (2019) no.5, 59001
doi:10.1209/0295-5075/128/59001
[arXiv:1911.04482 [gr-qc]].

\bibitem{Nojiri:2020wmh}
S.~Nojiri, S.~D.~Odintsov, V.~K.~Oikonomou and T.~Paul,
Phys. Rev. D \textbf{102} (2020) no.2, 023540
doi:10.1103/PhysRevD.102.023540
[arXiv:2007.06829 [gr-qc]].

\bibitem{Nojiri:2019yzg}
S.~Nojiri, S.~D.~Odintsov and E.~N.~Saridakis,
Nucl. Phys. B \textbf{949} (2019), 114790
doi:10.1016/j.nuclphysb.2019.114790
[arXiv:1908.00389 [gr-qc]].

\bibitem{Brevik:2019mah}
I.~Brevik and A.~Timoshkin,
doi:10.1142/S0219887820500231
[arXiv:1911.09519 [gr-qc]].

\bibitem{Coriano:2019eif}
C.~Corianno and P.~H.~Frampton,
Mod. Phys. Lett. A \textbf{35} (2019) no.02, 1950355
doi:10.1142/S0217732319503553
[arXiv:1906.10090 [gr-qc]].


\bibitem{Elizalde:2020zcb}
E.~Elizalde, S.~D.~Odintsov, V.~K.~Oikonomou and T.~Paul,
Nucl. Phys. B \textbf{954} (2020), 114984
doi:10.1016/j.nuclphysb.2020.114984
[arXiv:2003.04264 [gr-qc]].


\bibitem{Odintsov:2020zct}
S.~D.~Odintsov, V.~K.~Oikonomou and T.~Paul,
Class. Quant. Grav. \textbf{37} (2020) no.23, 235005
doi:10.1088/1361-6382/abbc47
[arXiv:2009.09947 [gr-qc]].


\bibitem{Tsallis:1987eu}
C.~Tsallis,
J.\ Statist.\ Phys.\ {\bf 52} (1988) 479.

\bibitem{Lyra:1998wz} 
M.~L.~Lyra and C.~Tsallis,
Phys.\ Rev.\ Lett.\ {\bf 80}, 53 (1998).

\bibitem{Wilk:1999dr} 
G.~Wilk and Z.~Wlodarczyk,
Phys.\ Rev.\ Lett.\ {\bf 84}, 2770 (2000)
[hep-ph/9908459].

\bibitem{Tsallis:2012js} 
C.~Tsallis and L.~J.~L.~Cirto,
Eur.\ Phys.\ J.\ C {\bf 73}, 2487 (2013)
[arXiv:1202.2154 [cond-mat.stat-mech]].

\bibitem{Komatsu:2013qia} 
N.~Komatsu and S.~Kimura,
Phys.\ Rev.\ D {\bf 88}, 083534 (2013)
[arXiv:1307.5949 [astro-ph.CO]].

\bibitem{Nunes:2014jra} 
E.~M.~Barboza, Jr., R.~d.~C.~Nunes, E.~M.~C.~Abreu and J.~Ananias Neto,
Physica A {\bf 436}, 301 (2015)
[arXiv:1403.5706 [gr-qc]].

\bibitem{Lymperis:2018iuz} 
A.~Lymperis and E.~N.~Saridakis,
Eur.\ Phys.\ J.\ C {\bf 78}, no. 12, 993 (2018)
doi:10.1140/epjc/s10052-018-6480-y
[arXiv:1806.04614 [gr-qc]].

\bibitem{Saridakis:2018unr} 
E.~N.~Saridakis, K.~Bamba, R.~Myrzakulov and F.~K.~Anagnostopoulos,
JCAP {\bf 1812}, no. 12, 012 (2018)
[arXiv:1806.01301 [gr-qc]].

\bibitem{Sheykhi:2018dpn} 
A.~Sheykhi,
Phys.\ Lett.\ B {\bf 785}, 118 (2018)
[arXiv:1806.03996 [gr-qc]].

\bibitem{Artymowski:2018pyg} 
M.~Artymowski and J.~Mielczarek,
arXiv:1806.03924 [gr-qc].

\bibitem{Abreu:2017hiy} 
E.~M.~C.~Abreu, J.~A.~Neto, A.~C.~R.~Mendes and A.~Bonilla,
EPL {\bf 121}, no. 4, 45002 (2018)
[arXiv:1711.06513 [gr-qc]].

\bibitem{Jawad:2018frc} 
A.~Jawad and A.~Iqbal,
doi:10.1142/S021988781850130X

\bibitem{Zadeh:2018wub} 
M.~Abdollahi Zadeh, A.~Sheykhi and H.~Moradpour,
arXiv:1810.12104 [physics.gen-ph].

\bibitem{daSilva:2018ehn} 
W.~J.~C.~da Silva and R.~Silva,
arXiv:1810.03759 [astro-ph.CO].

\bibitem{Biro:2013cra}
T.~S.~Bir\'o and V.~G.~Czinner,
Phys. Lett. B \textbf{726} (2013), 861-865
doi:10.1016/j.physletb.2013.09.032
[arXiv:1309.4261 [gr-qc]].

\bibitem{Czinner:2015eyk}
V.~G.~Czinner and H.~Iguchi,
Phys. Lett. B \textbf{752} (2016), 306-310
doi:10.1016/j.physletb.2015.11.061
[arXiv:1511.06963 [gr-qc]].

\bibitem{Komatsu:2016vof}
N.~Komatsu,
Eur. Phys. J. C \textbf{77} (2017) no.4, 229
doi:10.1140/epjc/s10052-017-4800-2
[arXiv:1611.04084 [gr-qc]].

\bibitem{Moradpour:2017ycq}
H.~Moradpour, A.~Bonilla, E.~M.~C.~Abreu and J.~A.~Neto,
Phys. Rev. D \textbf{96} (2017) no.12, 123504
doi:10.1103/PhysRevD.96.123504
[arXiv:1711.08338 [physics.gen-ph]].

\bibitem{Moradpour:2018ima}
H.~Moradpour, A.~Sheykhi, C.~Corda and I.~G.~Salako,
Phys. Lett. B \textbf{783} (2018), 82-85
doi:10.1016/j.physletb.2018.06.040
[arXiv:1711.10336 [physics.gen-ph]].

\bibitem{Moradpour:2018ivi}
H.~Moradpour, S.~A.~Moosavi, I.~P.~Lobo, J.~P.~Morais Gra\c{c}a, A.~Jawad and I.~G.~Salako,
Eur. Phys. J. C \textbf{78} (2018) no.10, 829
doi:10.1140/epjc/s10052-018-6309-8
[arXiv:1803.02195 [physics.gen-ph]].

\bibitem{Jahromi:2018xxh}
A.~Sayahian Jahromi, S.~A.~Moosavi, H.~Moradpour, J.~P.~Morais Gra\c{c}a, I.~P.~Lobo, I.~G.~Salako and A.~Jawad,
Phys. Lett. B \textbf{780} (2018), 21-24
doi:10.1016/j.physletb.2018.02.052
[arXiv:1802.07722 [gr-qc]].


\bibitem{Masi}
M.~Masi, 
Phys. Let. A \textbf{338} (2005), 217-224 
doi: 10.1016/j.physleta.2005.01.094
[arXiv:cond-mat/0505107 [cond-mat.stat-mech]].




\bibitem{Halliwell:1986ja}
J.~J.~Halliwell,
Phys. Lett. B \textbf{185} (1987), 341
doi:10.1016/0370-2693(87)91011-2

\bibitem{Barreiro:1999zs}
T.~Barreiro, E.~J.~Copeland and N.~J.~Nunes,
Phys. Rev. D \textbf{61} (2000), 127301
doi:10.1103/PhysRevD.61.127301
[arXiv:astro-ph/9910214 [astro-ph]].

\bibitem{Rubano:2001su}
C.~Rubano and P.~Scudellaro,
Gen. Rel. Grav. \textbf{34} (2002), 307-328
doi:10.1023/A:1015395512123
[arXiv:astro-ph/0103335 [astro-ph]].

\bibitem{Sangwan:2018zpz}
A.~Sangwan, A.~Tripathi and H.~K.~Jassal,
[arXiv:1804.09350 [astro-ph.CO]].

\bibitem{Adak:2013vwa}
  D.~Adak, A.~Ali and D.~Majumdar,
  Phys.\ Rev.\ D {\bf 88} (2013) no.2,  024007
  doi:10.1103/PhysRevD.88.024007
  [arXiv:1305.2330 [astro-ph.CO]].

\bibitem{delCampo:2013hka}
S.~del Campo, J.~C.~Fabris, R.~Herrera and W.~Zimdahl,
Phys. Rev. D \textbf{87} (2013) no.12, 123002
doi:10.1103/PhysRevD.87.123002
[arXiv:1303.3436 [astro-ph.CO]].

\bibitem{Granda:2008dk}
L.~N.~Granda and A.~Oliveros,
Phys. Lett. B \textbf{669} (2008), 275-277
doi:10.1016/j.physletb.2008.10.017
[arXiv:0810.3149 [gr-qc]].

\bibitem{Bekenstein:1974ax}
J.~D.~Bekenstein,
Phys. Rev. D \textbf{9} (1974), 3292-3300
doi:10.1103/PhysRevD.9.3292

\bibitem{Hawking:1974sw}
S.~W.~Hawking,
Commun. Math. Phys. \textbf{43} (1975), 199-220
[erratum: Commun. Math. Phys. \textbf{46} (1976), 206]
doi:10.1007/BF02345020


\bibitem{Jacobson:1995ab} 
T.~Jacobson,
Phys.\ Rev.\ Lett.\ {\bf 75}, 1260 (1995)
[gr-qc/9504004].

\bibitem{Padmanabhan:2003gd} 
T.~Padmanabhan,
Phys.\ Rept.\ {\bf 406}, 49 (2005)
[gr-qc/0311036].

\bibitem{Padmanabhan:2009vy} 
T.~Padmanabhan,
Rept.\ Prog.\ Phys.\ {\bf 73}, 046901 (2010)
[arXiv:0911.5004 [gr-qc]].

\bibitem{Cai:2005ra}
R.~G.~Cai and S.~P.~Kim,
JHEP \textbf{02} (2005), 050
doi:10.1088/1126-6708/2005/02/050
[arXiv:hep-th/0501055 [hep-th]].

\bibitem{Akbar:2006kj} 
M.~Akbar and R.~G.~Cai,
Phys.\ Rev.\ D {\bf 75}, 084003 (2007)
[hep-th/0609128].

\bibitem{Cai:2006rs} 
R.~G.~Cai and L.~M.~Cao,
Phys.\ Rev.\ D {\bf 75}, 064008 (2007)
[gr-qc/0611071].

\bibitem{Nojiri:2019itp}
S.~Nojiri, S.~D.~Odintsov, E.~N.~Saridakis and R.~Myrzakulov,
Nucl. Phys. B \textbf{950} (2020), 114850
doi:10.1016/j.nuclphysb.2019.114850
[arXiv:1911.03606 [gr-qc]].

\bibitem{Nojiri:2019skr}
S.~Nojiri, S.~D.~Odintsov and E.~N.~Saridakis,
Eur.\ Phys.\ J.\ C {\bf 79} (2019) no.3,  242
[arXiv:1903.03098 [gr-qc]].


\bibitem{Maeder:2021rqd}
A.~Maeder and V.~G.~Gueorguiev,
doi:10.1093/mnras/stab1102
[arXiv:2104.09314 [gr-qc]].

\end{thebibliography}
\end{document}